\documentclass[12pt, twocolumn]{openjournal}

\usepackage{xcolor}
\usepackage{textgreek}
\usepackage[utf8]{inputenc}
\usepackage[english]{babel}
\usepackage{hyperref}
\hypersetup{
    colorlinks=true,
    linkcolor=blue,
    filecolor=blue,      
    urlcolor=blue,
    citecolor=blue,
}
\usepackage{color,colortbl}
\usepackage{tensind}
\tensordelimiter{?}
\DeclareGraphicsExtensions{.bmp,.png,.jpg,.pdf}
\usepackage{verbatim}
\usepackage[normalem]{ulem}
\usepackage{orcidlink}
\usepackage{soul}
\usepackage{amsmath}

\def\gsim{\gtrsim}
\def\lsim{\lesssim}

\def\postsubmit#1{#1}

\graphicspath{ {./figs/} }

\begin{document}

\title{On the minimum number of radiation field parameters to specify gas cooling and heating functions}

\author{David Robinson\orcidlink{0000-0002-3751-6145}}
\email{dbrobins@umich.edu}
\affiliation{Department of Physics; University of Michigan, Ann Arbor, MI 48109, USA}
\affiliation{Leinweber Center for Theoretical Physics; University of Michigan, Ann Arbor, MI 48109, USA}

\author{Camille Avestruz\orcidlink{0000-0001-8868-0810}}
\affiliation{Department of Physics; University of Michigan, Ann Arbor, MI 48109, USA}
\affiliation{Leinweber Center for Theoretical Physics; University of Michigan, Ann Arbor, MI 48109, USA}

\author{Nickolay Y. Gnedin}
\affiliation{Theory Division; 
Fermi National Accelerator Laboratory;
Batavia, IL 60510, USA}
\affiliation{Kavli Institute for Cosmological Physics;
The University of Chicago;
Chicago, IL 60637, USA}
\affiliation{Department of Astronomy \& Astrophysics; 
The University of Chicago; 
Chicago, IL 60637, USA}

\begin{abstract}
    Fast and accurate approximations of gas cooling and heating functions are needed for hydrodynamic galaxy simulations. We use machine learning to analyze atomic gas cooling and heating functions computed by Cloudy in the presence of a generalized incident local radiation field.  We characterize the radiation field through binned radiation field intensities instead of the photoionization rates used in our previous work. We find a set of 6 energy bins whose intensities exhibit relatively low correlation. We use these bins as features to train machine learning models to predict Cloudy cooling and heating functions at fixed metallicity. We compare the relative \postsubmit{SHapley Additive exPlanation (SHAP) value} importance of the features.  From the SHAP analysis, we identify a feature subset of 3 energy bins ($0.5-1, 1-4$, and $13-16 \, \mathrm{Ry}$) with the largest importance and train additional models on this subset. We compare the mean squared errors and distribution of errors on both the entire training data table and a randomly selected 20\% test set withheld from model training.  The machine learning models trained with 3 and 6 bins, as well as 3 and 4 photoionization rates, have comparable accuracy everywhere\postsubmit{, with errors $\gtrsim 10$ times smaller than for the interpolation table of \citet{gnedin_hollon12}}. We conclude that 3 energy bins (or 3 analogous photoionization rates: molecular hydrogen photodissociation, neutral hydrogen \ion{H}{1}, and fully ionized carbon \ion{C}{6}) are sufficient to characterize the dependence of the gas cooling and heating functions on our assumed incident radiation field model.
\end{abstract}

\begin{keywords}
    {galaxies:formation, methods:numerical, cosmology:miscellaneous}
\end{keywords}

\maketitle

\section{Introduction}
\label{sec:intro}

The various feedback processes through which stars and active galactic nuclei deposit energy, momentum, and gas enriched with metals and dust into the circumgalactic medium (CGM) and back into the interstellar medium (ISM) from which they formed, are crucial parts of galaxy formation and evolution. One particular mode of feedback is direct photoionization and photoheating of gas by the local interstellar radiation field (ISRF), which modifies the thermodynamics of the gas \citep[see, e.g., the reviews of][]{benson10, naab_ostriker17}.

Cooling and heating functions describe how the energy density of a cloud of gas changes in time due to radiative processes \citep[e.g.][]{cox_tucker69, sutherland_dopita93}, and so will be affected by photoionization and photoheating feedback.  Cooling and heating functions are critical to understanding galaxy formation and evolution \citep[e.g., the reviews of][]{benson10, naab_ostriker17}.  The energy density of a gas cloud can also change due to mechanical expansion or compression.  In an idealized scenario without star formation, the evolution of the size of the cloud is determined by the balance between (net) radiative cooling and gravitational compression \citep[e.g.][]{lee24}.  This same balance also occurs in the process of star formation.  Including the effects of photoionization feedback in the cooling function has been shown to change the stellar mass formed in a simulated Milky Way-like galaxy \citep{kannan14}.

The incident radiation field in galaxies, which affects the cooling and heating functions, can have contributions from a variety of sources. External galaxies and quasars combine to produce a mildly undulating  (but time-varying) ultraviolet (UV) background of ionizing radiation.  Theoretical models for (the spatial average of) this background include \citet{faucher_giguere09} and \citet{hardt_madau12}.  The interstellar radiation field within the Milky Way depends on location through the spatial positions of sources (including stars) and absorbers such as gas and dust \citep{popescu17}.

In a cosmological or galaxy evolution simulation with radiative transfer, we can calculate the radiation field in various energy bands. Some recent works have focused on modeling the radiation field in simulations across a wide range of wavelengths \citep[e.g.,][]{romero23}, or in specific wavelength ranges such as polycyclic aromatic hydrocarbon emission in the infrared \citep[e.g.,][]{narayanan23}; the ultraviolet Lyman and Werner bands for the photodissociation of molecular hydrogen \citep[e.g.,][]{incatasciato23}; and ultraviolet bands containing the photoionization thresholds for neutral hydrogen (\ion{H}{1}), neutral helium (\ion{He}{1}), and singly ionized helium (\ion{He}{2}), among others \citep[e.g.,][]{kannan20, baumschlager23}.  \citet{baumschlager23} uses narrow energy bins of width $0.1 \, \mathrm{eV}$ at the edges of the \ion{H}{1}, \ion{He}{1}, and \ion{He}{2}-ionizing bins, in order to model the radiation field spectrum in each of those bins as a power law (as opposed to a constant).  

Given an arbitrary incident radiation field, we can evaluate a network of differential equations for the creation and destruction of each relevant ion by chemical reactions, collisions, and absorption and emission of radiation.  The net emission and absorption of photons at each relevant energy determine the cooling and heating functions.  This is implemented in both photoionization codes like Cloudy \citep{ferland98}, and chemical networks such as \texttt{GRACKLE} \citep{smith17} and \texttt{KROME} \citep{grassi14}.  These networks can have varying degrees of complexity (i.e. numbers of ions and reactions incorporated). Including more ions makes the calculations more exact at the cost of increased computational complexity.  Including fewer ions decreases the accuracy, but can make the networks fast enough to utilize in a hydrodynamic simulation \citep[e.g.][]{grassi14, smith17}.  Another approach to simplifying the computational complexity of large chemical networks is to emulate the network with a machine learning algorithm \citep[e.g.][]{branca_pallottini22, branca_pallottini24}.

In general, approximating cooling and heating functions subject to an incident radiation field requires making some assumptions to simplify the radiation field to a finite number of parameters. The simplest possible radiation field, $J_\nu = 0$, has no parameters.  In this limit, known as collisional ionization equilibrium (CIE), no ionizing radiation is incident on the gas, and the cooling and heating functions depend only on properties of the gas \citep[density, temperature, and metallicity, e.g.,][]{cox_tucker69, sutherland_dopita93}. Since the extragalactic UV background radiation depends only on redshift, the cooling and heating functions of a gas irradiated by this background depend on redshift, in addition to gas properties \citep[e.g.,][]{kravtsov03, wiersma09}.  \citet{gnedin_hollon12} assume a 4-parameter model for the local radiation field: an arbitrary mixture of a synthesized stellar spectrum and a quasar-like power law, optionally attenuated by an additional optical depth due to the local absorption by \ion{H}{1} and \ion{He}{1} (this spectrum is discussed in more detail in Section~\ref{method:training_data} and \ref{methods:rf_bins}), and tabulate the resulting cooling and heating functions using photoionization rates calculated from that spectrum.  Another possibility is to use the average intensity of the radiation field in various energy bins as the parameters, as is done in \citet{branca_pallottini24}. \postsubmit{Rather than interpolating on a grid of parameter values, \citet{galligan19} uses gas and radiation field properties from a zoom-in galaxy simulation with radiative transfer to run Cloudy models to calculate the cooling and heating functions, and trains a machine learning algorithm to interpolate the results.}

In this paper, we utilize \postsubmit{a} machine learning framework to approximate cooling and heating functions that we developed in \citet{robinson24}. 
We parameterize the radiation field with its average intensity in various energy bins instead of the photoionization rates used in \citet{robinson24}. Using radiation field bins is useful because many radiative transfer simulations calculate the radiation field in such bins (see above).  These bins also summarize the radiation field in a different way than photoionization rates.  In particular, they do not include the frequency dependence of photoionization cross sections.  So, comparing the performance of machine learning models trained using binned radiation field intensities with those trained using photoionization rates allows us to gain more insight into what aspects of the incident radiation field are most important to cooling and heating functions.

Following the approach of \citet{robinson24}, we use a training table from \citet{gnedin_hollon12}, which consists of cooling and heating functions calculated with Cloudy across a range of temperatures, densities, metallicities, and 4 radiation field parameters (see Table~\ref{table:grid_params}).  We apply the machine learning algorithm eXtreme Gradient Boosting \citep[XGBoost,][]{chen_guestrin16} to predict the cooling and heating functions \textit{at fixed metallicity}.  XGBoost is an example of a gradient-boosted tree algorithm, where an ensemble of trees are trained \textit{sequentially}.  Each new tree is trained to predict the difference between the true value and the prediction from the previously-trained trees.  The final prediction is then the sum of the predictions from each tree.  At each tree node, the branches split based on the value of one feature \citep{chen_guestrin16}.  Gradient-boosted tree algorithms have been shown to outperform deep-learning methods on problems where the training data is in tabular form \citep{shwartz-ziv_armon21, grinztajn22}, such as the training data used here.  

XGBoost also interfaces easily with SHapley Additive exPlanation (SHAP) values \citep{lundberg_lee17, lundberg18, lundberg20}, which we use to study the relative importance of various radiation field bins.  For any given model prediction, the SHAP value for each feature is the difference between the actual model prediction and an expected prediction ignoring the feature in question.  This definition ensures that the SHAP values for every feature add up to the exact point prediction in question, and gives 0 for any feature that does not affect the model prediction.  But, training models on all subsets of features is impractical for models with a significant number of features \citep{lundberg_lee17}, leading to the development of tools to \textit{approximate} SHAP values for various kinds of models \citep{lundberg18, lundberg20}.  

In this paper we present our methodology in Section~\ref{sec:methods}, including our training data, how we choose radiation field bins, our machine learning pipeline, and how we use SHAP values to compare the importance of various model features.  Then, in Section~\ref{sec:results}, we evaluate the performance of the various machine learning models trained in this paper.  We conclude with a summary and discussion in Section~\ref{sec:concl}.

\section{Methodology}
\label{sec:methods}

\subsection{The training grid} \label{method:training_data}
Similarly to \citet{robinson24}, we train our machine learning models to predict atomic gas cooling and heating functions computed with the photoionization code Cloudy \citep{ferland98}, using the same data table as \citet{gnedin_hollon12}. We define the gas heating function $\Gamma$ and cooling function $\Lambda$ through the equation
\begin{equation}
    \left. \frac{\mathrm{d}U}{\mathrm{d}t} \right\vert_\mathrm{rad} = n_b^2 \left[ \Gamma(T, n_b, Z, J_\nu) - \Lambda(T, n_b, Z, J_\nu) \right],
    \label{eq:chf_def}
\end{equation}
where $U$ is the thermal energy density of the gas, $Z$ is its metallicity, $n_b = n_\mathrm{H} + 4n_\mathrm{He} + \ldots$ is its baryon number density, and $J_\nu$ is the specific intensity of the incident radiation field.  The factor of $n_b^2$ accounts for the expected density dependence in the limit of collisional ionization equilibrium, and with only two-body collisions.  However, in the more general case considered here, the cooling and heating functions will depend on $n_b$ (and also $Z$ and $J_\nu$).

We assume a radiation field model with specific intensity given by
\begin{equation}
    J_\nu = J_0 \left[ \frac{1}{1+f_Q} s_\nu + \frac{f_Q}{1+f_Q}x^{-\alpha} \right]e^{-\tau_\nu},
    \label{eq:rad_field}
\end{equation}
where the overall amplitude $J_0$ is a free parameter, and $x = h\nu / (1 \, \mathrm{Ry})$ is the photon energy in Rydbergs ($1 \, \mathrm{Ry} = 13.6 \, \mathrm{eV}$, the ionization threshold for neutral hydrogen \ion{H}{1}).  The first term in the square brackets is a stellar spectrum fit from the spectral synthesis library Starburst99 \citep{starburst99}:
\begin{equation}
    s_\nu = \frac{1}{5.5} 
    \begin{cases}
    5.5 & x < 1, \\
    x^{-1.8} & 1 < x < 2.5, \\
    0.4x^{-1.8} & 2.5 < x < 4, \\
    2 \times 10^{-3} \frac{x^3}{e^{x/1.4} - 1} & x > 4. \\
    \end{cases}
    \label{eq:stellar_spec}
\end{equation}
The second term is a quasar-like power law with index $\alpha$ (another free parameter).  The free parameter $f_Q$ is the strength of the quasar-like component relative to the stellar spectrum at $x = 1$. The entire spectrum is also attenuated by neutral hydrogen and helium (\ion{He}{1}) absorption with optical depth
\begin{equation}
    \tau_\nu = \frac{\tau_0}{\sigma_{\mathrm{HI, 0}}} \left[0.76\sigma_\mathrm{HI}(\nu) + 0.06\sigma_\mathrm{HeI}(\nu) \right],
    \label{eq:opt_depth}
\end{equation}
where $\sigma_{j}(\nu)$ is the photoionization cross section as a function of frequency for ion $j$, computed using the fits from \citet{verner96}, and $\sigma_{j, 0}$ is that same cross section evaluated at the threshold frequency. Finally, $\tau_0$ is a dimensionless free parameter which scales the overall optical depth.  That is, it describes the gas column density attenuating the radiation field.

The training data consists of Cloudy models evaluated on a grid of values for the gas temperature $T$, hydrogen number density $n_H$, and metallicity $Z$, as well as the radiation field parameters $J_0, f_Q, \tau_0, $ and $\alpha$.  We describe the values used in Table~\ref{table:grid_params}. \postsubmit{For $Z/\mathrm{Z}_\odot \lesssim 0.01$, the cooling and heating functions do not depend on metallicity \citep[metal-line cooling is unimportant,][]{boehringer89}. Hence, we choose $Z/\mathrm{Z}_\odot = 0.1$ as the lowest non-zero metallicity value.}\footnote{The $Z/\mathrm{Z}_\odot=0$ case actually has $Z/\mathrm{Z}_\odot = 10^{-4}$, but that value is small enough to be indistinguishable from the primordial gas for our purposes.} \postsubmit{We also require $Z/\mathrm{Z}_\odot \leq 3$ as higher metallicity values are unlikely to occur in the ISM or CGM.} This results in $81 \times 13 \times 5 \times 25 \times 9 \times 9 \times 7 \approx 7.5 \times 10^{7}$ grid points.

\begingroup 
    \setlength{\tabcolsep}{6pt} 
    \renewcommand{\arraystretch}{1.5} 
    \begin{table}
        \centering
        \begin{tabular}{ r l }  
            Parameter & Values \\
            \hline \hline     
            $\log{(T/\mathrm{K})}$ & $1, 1.1, 1.2, \ldots, 9$ \\
            $\log{(n_\mathrm{H}/\mathrm{cm}^{-3})}$ & $-6, -5, -4, \ldots, 6$ \\   
            $Z/\mathrm{Z}_\odot$ & $0, 0.1, 0.3, 1, 3$ \\
            \hline 
            $\log{(J_0 \, \mathrm{cm}^{-3}/n_b/J_\mathrm{MW})}$ & $-5, -4.5, -4, \ldots, 7$ \\
            $\log{(f_Q)}$ & $-3, -2.5, -2, \ldots, 1$ \\
            $\log{(\tau_0)}$ & $-1, -0.5, 0, \ldots, 3$ \\
            $\alpha$ & $0, 0.5, 1, \ldots, 3$ \\
            \hline \hline
        \end{tabular}  
        \caption{Gas and radiation field parameters used to train XGBoost models of Cloudy-computed cooling and heating functions. We normalize the radiation field amplitude $J_0$ by $J_\mathrm{MW} = 10^{6} \, \mathrm{photons} \, \mathrm{cm}^{-2} \, \mathrm{s}^{-1} \, \mathrm{ster}^{-1} \, \mathrm{eV}^{-1}$.}
        \label{table:grid_params}
    \end{table}
\endgroup

\subsection{Radiation field sampling} \label{methods:rf_bins}
Note that if a radiation field $J_\nu$ does not have precisely the form given by Equation~(\ref{eq:rad_field}), the field may not have well-defined values of the parameters $J_0, f_Q, \tau_0,$ and $\alpha$.  To represent the radiation field, we instead compress the spectrum $J_\nu$ into a few parameters $r_j$ by integrating over the frequency domain with some weight functions $w_j(\nu)$, 
\begin{equation}
    \label{eq:rf_int}
    r_j = \int_{0}^{\infty} J_\nu w_j(\nu) \, \mathrm{d}\nu.
\end{equation}
In a previous study \citep{robinson24}, we used weights $w_j(\nu)=4\pi\sigma_j(\nu)/(h\nu)$, where $\sigma_j(\nu)$ is the frequency-dependent cross section for photoionization of the ion $j$.  With this choice, Equation~(\ref{eq:rf_int}) yields the photoionization rates
\begin{equation}
    \label{eq:q_j_def}
    P_j = c\int_0^\infty \sigma_j(\nu)n_\nu \, \mathrm{d}\nu,
\end{equation}
where $n_\nu = 4\pi J_\nu/(ch\nu)$ is the number density of photons at frequency $\nu$.  

However, we could in principle choose \textit{any} sufficiently large set of weight function $w_j(\nu)$, and we are not aware of a rigorous way to find an optimal choice for such a set. In the absence of a way to define optimal weighting functions, we choose to use the simple case of constant weights $w(\nu)={\rm const}$ within a fixed interval $\nu_a<\nu<\nu_b$.  More specifically, we compute the average intensity in a frequency bin:
\begin{equation}
    \langle J_{\nu_a - \nu_b} \rangle = \frac{\int_{\nu_a}^{\nu_b}J_\nu \, \mathrm{d}\nu}{\nu_b - \nu_a}.
    \label{eq:binned_rf}
\end{equation}
for several bins as our parameters for representing the radiation field.

This definition still leaves a choice of bin edges $\nu_a$ and $\nu_b$. To choose these bin edges, we consider the correlations between different $\langle J_{\nu_a - \nu_b} \rangle$ values. Plugging Equation~(\ref{eq:rad_field}) into Equation~(\ref{eq:binned_rf}) indicates that $\langle J_{\nu_a - \nu_b} \rangle$ scales linearly with the overall amplitude $J_0$ for all bins $\nu_a - \nu_b$.  To account for this, we choose the bin $0.5 - 1 \, \mathrm{Ry}$, just below the ionization threshold for \ion{H}{1}, and divide all other $\langle J_{\nu_a - \nu_b} \rangle$ by $\langle J_{0.5-1 \, \mathrm{Ry}} \rangle$ to remove the dependence on $J_0$. This is directly analogous to our use of the photodissociation rate of molecular hydrogen $P_\mathrm{LW}$ to scale all other photoionization rates $P_j$ in \citet{robinson24}.  The photons that dissociate molecular hydrogen have energies of about $0.82-1 \, \mathrm{Ry}$ \citep{draine_bertoldi96}, lying within this $0.5-1 \, \mathrm{Ry}$ bin.

We are now faced with the question of how to choose bin edges $\nu_a$ and $\nu_b$ for the values $\langle J_{\nu_a - \nu_b} \rangle/\langle J_{0.5-1 \, \mathrm{Ry}} \rangle$. Two options for selecting bin edges are linear spacing and logarithmic spacing.  Linear spacing produces bins of constant width, while logarithmic spacing creates bins \postsubmit{that} increase in width exponentially with increasing frequency. Our choices here are guided by the correlations between bins and heuristic physical considerations.  For example, the ionization energies for helium are approximately $1.8 \, \mathrm{Ry}$ for \ion{He}{1} and $4 \, \mathrm{Ry}$ for doubly ionized helium \ion{He}{2} \citep{verner96}.  We select a $1-4 \, \mathrm{Ry}$ bin to include the ionization thresholds for \ion{H}{1} and \ion{He}{1}, but exclude \ion{He}{2} ionizing radiation.\footnote{See appendix~\ref{ap:log_spaced_bins} for a discussion of using two energy bins in the $1-4 \, \mathrm{Ry}$ range.}

\begin{figure*}
    \centering
    \includegraphics[width = \textwidth]{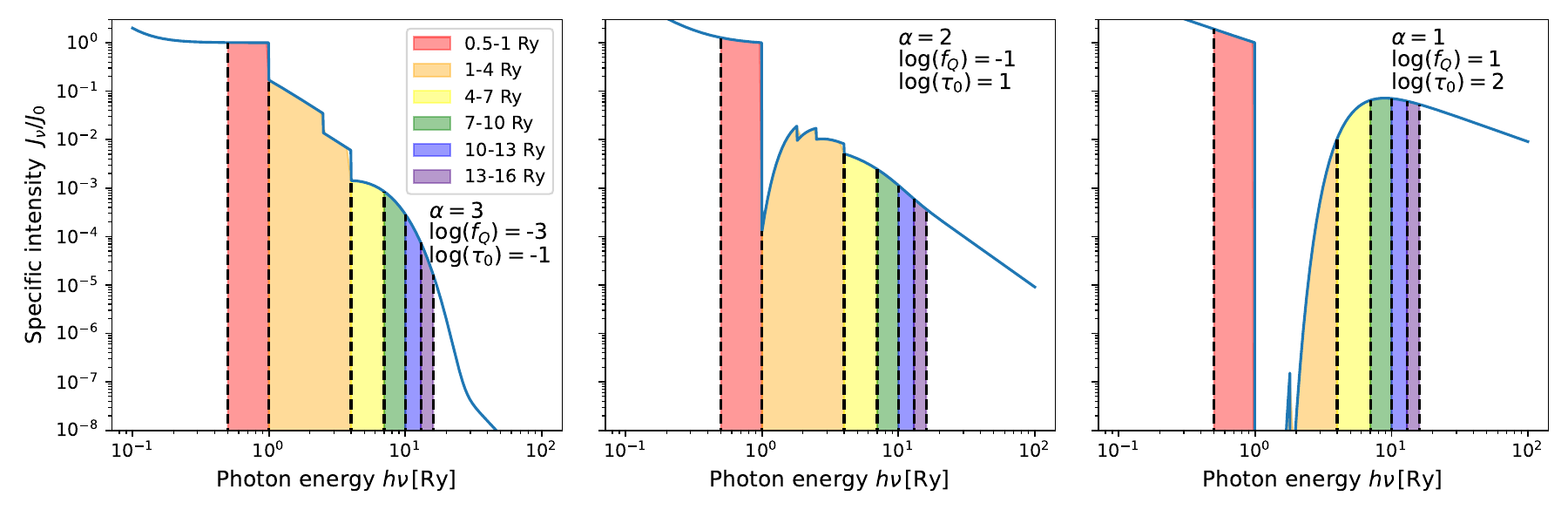}
    \caption{Examples of the radiation field $J_\nu$ (normalized by its overall amplitude $J_0$) defined by Equation~(\ref{eq:rad_field}) for different choices of $\alpha, f_Q,$ and $\tau_0$.  The dashed vertical lines indicate our choice of bin edges, at $0.5, 1, 4, 7, 10, 13,$ and $16 \, \mathrm{Ry}$. 
    }
    \label{fig:rf_examples}
\end{figure*}

To shed more light on the correlations between various $\langle J_{\nu_a - \nu_b} \rangle/\langle J_{0.5-1 \, \mathrm{Ry}} \rangle$ values, we consider how the radiation field $J_\nu$ given by 
Equation~(\ref{eq:rad_field}) changes with different choices of $f_Q$, $\tau_0,$ and $\alpha$.  Three examples are shown in Fig.~\ref{fig:rf_examples}, along with our choice of bin edges discussed below.
Comparing the three panels of Fig.~\ref{fig:rf_examples} shows that $J_\nu/J_0$ is strongly dependent on $\alpha$ at energies above about $16 \, \mathrm{Ry}$, but has only weak dependence on $f_Q$ and $\tau_0$.  So, we should expect all $\langle J_{\nu_a - \nu_b} \rangle/\langle J_{0.5-1 \, \mathrm{Ry}} \rangle$ to be highly correlated for $\nu_a \gtrsim 16 \, \mathrm{Ry}$. The correlations between bins with logarithmically spaced edges are explored further in Appendix~\ref{ap:log_spaced_bins}, but we find linear bins to work better.

\begin{figure*}
    \centering
    \includegraphics[width = \textwidth]{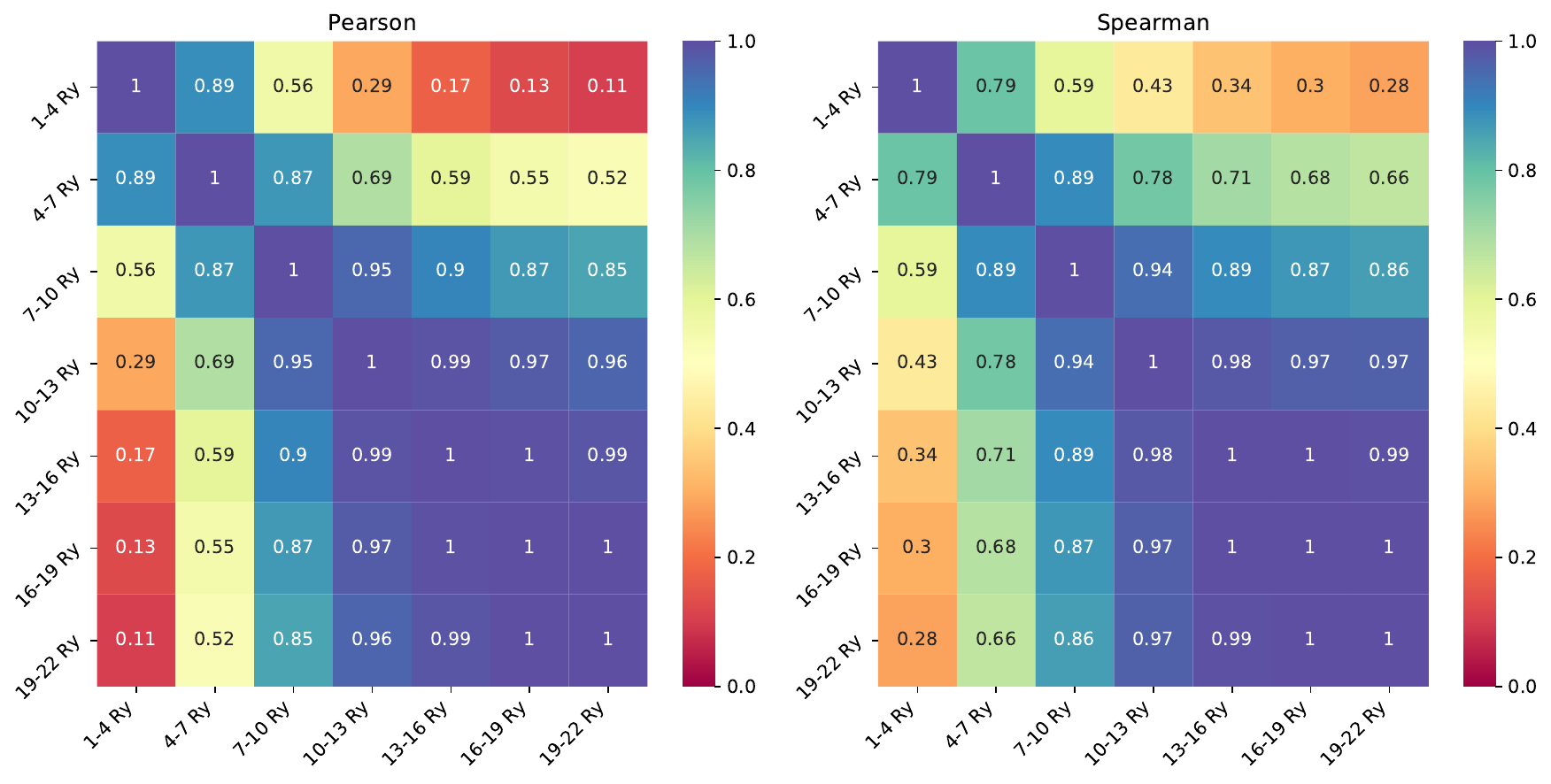}
    \caption{The absolute value of the Pearson (left panel) and Spearman (right panel) correlation coefficients between binned radiation field intensities $\log{(\langle J_{\nu_a - \nu_b} \rangle/\langle J_{0.5-1 \, \mathrm{Ry}} \rangle)}$ for the values in Table~\ref{table:grid_params}, with linear bin spacing.  Given the high correlation between the highest three energy bins, we exclude the two highest bins ($16-19 \, \mathrm{Ry}$ and $19-22 \, \mathrm{Ry}$) from our initial model training.}
    \label{fig:bin_corr}
\end{figure*}

The Pearson and Spearman correlation matrices of $\log{(\langle J_{\nu_a - \nu_b} \rangle/\langle J_{0.5-1 \, \mathrm{Ry}} \rangle)}$ for bins of constant width $h\nu_b - h\nu_a = 3 \, \mathrm{Ry}$, up to $h\nu_b = 22 \, \mathrm{Ry}$ are shown in Fig.~\ref{fig:bin_corr}.  The Spearman correlation only considers the \textit{ranks} of points.  Thus, these correlations are unaffected by taking the logarithm of the scaled intensities (as the logarithm is a monotonically increasing function).  In contrast, the Pearson correlation is sensitive to the values of individual points, not just their ranks. The bins with energies above $16 \, \mathrm{Ry}$ are very highly correlated, with both Pearson and Spearman correlations of at least $0.99$.  We explore energies up to $2048 \, \mathrm{Ry}$ (with logarithmic bin spacing) in Appendix~\ref{ap:log_spaced_bins}, but, as expected from Fig.~\ref{fig:rf_examples}, all bins above $16 \, \mathrm{Ry}$ are highly correlated with each other. The correlations are more modest for energies below $13 \, \mathrm{Ry}$.  

For any $f_Q > 0$, the radiation field spectrum given by Equation~(\ref{eq:rad_field}) becomes dominated by the quasar-like power law $x^{-\alpha}$ at sufficiently high photon energies $x$. Increasing $f_Q$ (moving left to right in Fig.~\ref{fig:rf_examples}) decreases the energy at which this power law becomes dominant.  Since the optical depth $\tau_\nu$ described in Equation~(\ref{eq:opt_depth}) includes absorption by \ion{H}{1} and \ion{He}{1}, the primary effect of increasing the optical depth scaling $\tau_0$ (moving left to right in Fig.~\ref{fig:rf_examples}) is to reduce $J_\nu$ in the $1-4 \, \mathrm{Ry}$ bin containing the ionization energies for \ion{H}{1} and \ion{He}{1}.  However, sufficiently large values of $\tau_0$ also reduce the specific intensity in bins \textit{above} the photon energy $4 \, \mathrm{Ry}$.  This can be seen in the $4-7$ and $7-10 \, \mathrm{Ry}$ bins in the rightmost panel of Fig.~\ref{fig:rf_examples}.  So, increasing $\tau_0$ decreases the flux in every bin above $1 \, \mathrm{Ry}$, but decreases the flux in the $1-4 \, \mathrm{Ry}$ bin most strongly. Recall here that the Pearson correlation coefficient is sensitive to the values of each $\log{(\langle J_{\nu_a - \nu_b} \rangle/\langle J_{0.5-1 \, \mathrm{Ry}} \rangle)}$, while the Spearman correlation coefficient depends only on their ranks. This suggests that the optical depth $\tau_0$ should cause the Pearson correlation coefficient for $\log{(\langle J_{1-4 \mathrm{Ry}} \rangle/\langle J_{0.5-1 \, \mathrm{Ry}} \rangle)}$ and $\log{(\langle J_{\nu_a - \nu_b} \rangle/\langle J_{0.5-1 \, \mathrm{Ry}} \rangle)}$ with $\nu_a \geq 4 \, \mathrm{Ry}$ to be smaller than the corresponding Spearman correlations (ignoring the effects of varying $f_Q$ and $\alpha$). From Fig.~\ref{fig:bin_corr}, we see that this is true for all but the $4-7 \, \mathrm{Ry}$ bin.

For the remainder of this paper, we focus on the less-correlated values of $\log{(\langle J_{\nu_a - \nu_b} \rangle/\langle J_{0.5-1 \, \mathrm{Ry}} \rangle)}$ from Fig.~\ref{fig:bin_corr}: $1-4, 4-7, 7-10, 10-13$ and $13-16 \, \mathrm{Ry}$ (as well as $\langle J_{0.5-1 \, \mathrm{Ry}} \rangle$, which incorporates the overall amplitude $J_0$).

\subsection{Machine learning models}
\label{method:ml_methods}

\postsubmit{Gradient-boosted tree algorithms like XGBoost are known to outperform other machine learning methods for tabular training data like the the grid described in Table~\ref{table:grid_params} \citep{shwartz-ziv_armon21, grinztajn22}. XGBoost also provides a useful framework for exploring a variety of radiation field parameters. Because each individual tree in an XGBoost model only utilizes (at most) a fixed fraction of the available features, the data size and computational cost scale weakly with the number of features. Furthermore, XGBoost interfaces easily with the feature importance tools described in Section~\ref{method:shap}.}

We train the machine learning algorithm XGBoost \citep{chen_guestrin16} to predict $\log(\Gamma)$ and $\log(\Lambda)$ \textit{at fixed values of the metallicity} $Z$. This results in 10 distinct models (cooling and heating at $Z/\mathrm{Z}_\odot = \{0, 0.1, 0.3, 1, 3\}$). \postsubmit{As discussed in \citet{robinson24}, we train separate models at each metallicity because the 5 metallicity values in the training grid (see Table~\ref{table:grid_params}) are not enough samples for XGBoost to outperform a manual quadratic interpolation in metallicity. Indeed, \citet{robinson24} found that this metallicity interpolation is the main bottleneck for accurately predicting cooling and heating functions at \textit{arbitrary} metallicities.} The models are trained to optimize the mean squared error (MSE), defined as:
\begin{equation}
    \label{eq:MSE}
    \mathrm{MSE} = \langle [\log{\mathcal{F}_\mathrm{true}} - (\log{\mathcal{F}})_\mathrm{pred}]^2 \rangle = \langle (\Delta \log \mathcal{F})^2 \rangle,
\end{equation}
where $\mathcal{F} = \Gamma$ or $\Lambda$. 

\begingroup 
    \setlength{\tabcolsep}{4pt} 
    \renewcommand{\arraystretch}{1.5} 
    \begin{table*}
        \centering
        \begin{tabular}{rlllll}
            Model name & 6 bins & 3 bins & 2 bins & 4 rates & 3 rates \\
            \hline \hline 
            Temperature & $T/\mathrm{K}$ & $T/\mathrm{K}$ & $T/\mathrm{K}$ & $T/\mathrm{K}$ & $T/\mathrm{K}$ \\
            Density & $n_\mathrm{H}/\mathrm{cm}^{-3}$ & $n_\mathrm{H}/\mathrm{cm}^{-3}$ & $n_\mathrm{H}/\mathrm{cm}^{-3}$ & $n_\mathrm{H}/\mathrm{cm}^{-3}$ & $n_\mathrm{H}/\mathrm{cm}^{-3}$ \\
            \hline 
            Amplitude $J_0$ & $\langle J_{0.5-1 \, \mathrm{Ry}} \rangle \, \mathrm{cm}^{-3}/n_b/J_\mathrm{MW}$ & $\langle J_{0.5-1 \, \mathrm{Ry}} \rangle \, \mathrm{cm}^{-3}/n_b/J_\mathrm{MW}$ & $\langle J_{1-4 \, \mathrm{Ry}} \rangle \, \mathrm{cm}^{-3}/n_b/J_\mathrm{MW}$ & $P_\mathrm{LW} \, \mathrm{cm}^{-3} / \mathrm{s}^{-1}$ & $P_\mathrm{LW} \, \mathrm{cm}^{-3} / \mathrm{s}^{-1}$ \\
            Other features & $\langle J_{1-4 \, \mathrm{Ry}} \rangle / \langle J_{0.5 - 1 \, \mathrm{Ry}} \rangle$ & $\langle J_{1-4 \, \mathrm{Ry}} \rangle / \langle J_{0.5 - 1 \, \mathrm{Ry}} \rangle$ & $\langle J_{13-16 \, \mathrm{Ry}} \rangle / \langle J_{1-4 \, \mathrm{Ry}} \rangle$ & $P_\mathrm{HI}/P_\mathrm{LW}$ & $P_\mathrm{HI}/P_\mathrm{LW}$\\
            & $\langle J_{4-7 \, \mathrm{Ry}} \rangle / \langle J_{0.5 - 1 \, \mathrm{Ry}} \rangle$ & $\langle J_{13-16 \, \mathrm{Ry}} \rangle / \langle J_{0.5 - 1 \, \mathrm{Ry}} \rangle$ &  & $P_\mathrm{HeI}/P_\mathrm{LW}$ & $P_\mathrm{CVI}/P_\mathrm{LW}$ \\
            & $\langle J_{7-10 \, \mathrm{Ry}} \rangle / \langle J_{0.5 - 1 \, \mathrm{Ry}} \rangle$ &  & & $P_\mathrm{CVI}/P_\mathrm{LW}$ & \\
            & $\langle J_{10-13 \, \mathrm{Ry}} \rangle / \langle J_{0.5 - 1 \, \mathrm{Ry}} \rangle$ &  &  &  & \\
            & $\langle J_{13-16 \, \mathrm{Ry}} \rangle / \langle J_{0.5 - 1 \, \mathrm{Ry}} \rangle$ &  &  &  & \\
            \hline \hline
        \end{tabular}
        \caption{The features sets used for each XGBoost model considered in this paper.  Note that all features are the logarithm of the quantities shown here. Above the line are features describing gas properties, which are the same for each of our models. Below the line are features related to the incident radiation field.  The first row describes the radiation field features containing the overall amplitude of the radiation field $J_0$ (see Equation~(\ref{eq:rad_field})). The remaining rows show all other radiation field features, which are scaled to the rate or bin containing the overall amplitude.}
        \label{tab:features}
    \end{table*}
\endgroup

The 8 dimensionless, logarithmically scaled inputs we use are described in the `6 bins' column of Table~\ref{tab:features}. Even after taking the logarithm, these features vary over quite different ranges (e.g., see Table~\ref{table:grid_params}). To ameliorate this, we linearly rescale each feature to the range $[0,1]$ for the values in Table~\ref{table:grid_params}. 

Before training the `final' XGBoost models we use for our analysis below, we perform an initial hyperparameter validation step. \postsubmit{We} use a random subsample of 80\% of the input data (described in Section~\ref{method:training_data}) for model training. \postsubmit{We} subselect 10\% of this \postsubmit{training} subset (i.e. 8\% of the total data) \postsubmit{for hyperparameter validation and train models with the optimal hyperparameters on the remaining 72\%}.  The remaining 20\% is used as a test set for final model evaluation.  \postsubmit{These fractions were chosen to ensure that the training, hyperparameter validation, and test sets are all large enough to accurately sample the distributions of each feature and that the training set contains a majority of the available data.} We use the same data splitting for all the models discussed below. 
See Appendix~\ref{ap:hyperparams} for a brief description of the hyperparameter optimization procedure, which we fully lay out in \citet{robinson24}. 

Finally, we retrain models using the optimized hyperparameters (found as described in Appendix~\ref{ap:hyperparams}) on both the same 80\% training subset described above (so we can analyze model performance on a 20\% test set withheld from model training) and on the entire input data. The model evaluation on the 20\% test set illustrates how well our models generalize to data they have not yet seen, providing an unbiased measure of how well our models perform.  \postsubmit{Training models on the entire training grid (and testing on the same data)} provides a quantification of an overfitted result and a potential best-case scenario of how well our models might do with a larger training sample.

\subsection{Feature importance with SHAP values} \label{method:shap}
\postsubmit{To evaluate the impact each feature has on model predictions, we use SHapley Additive exPlanation (SHAP) values \citep{lundberg_lee17, lundberg18, lundberg20}. SHAP values can be calculated for each feature for any given point prediction of a machine learning model. SHAP values describe how the model prediction differs from what would be expected from ignoring the value of the feature in question \citep{lundberg_lee17}.}

\begin{figure*}
    \centering
    \includegraphics[width = 0.9\textwidth]{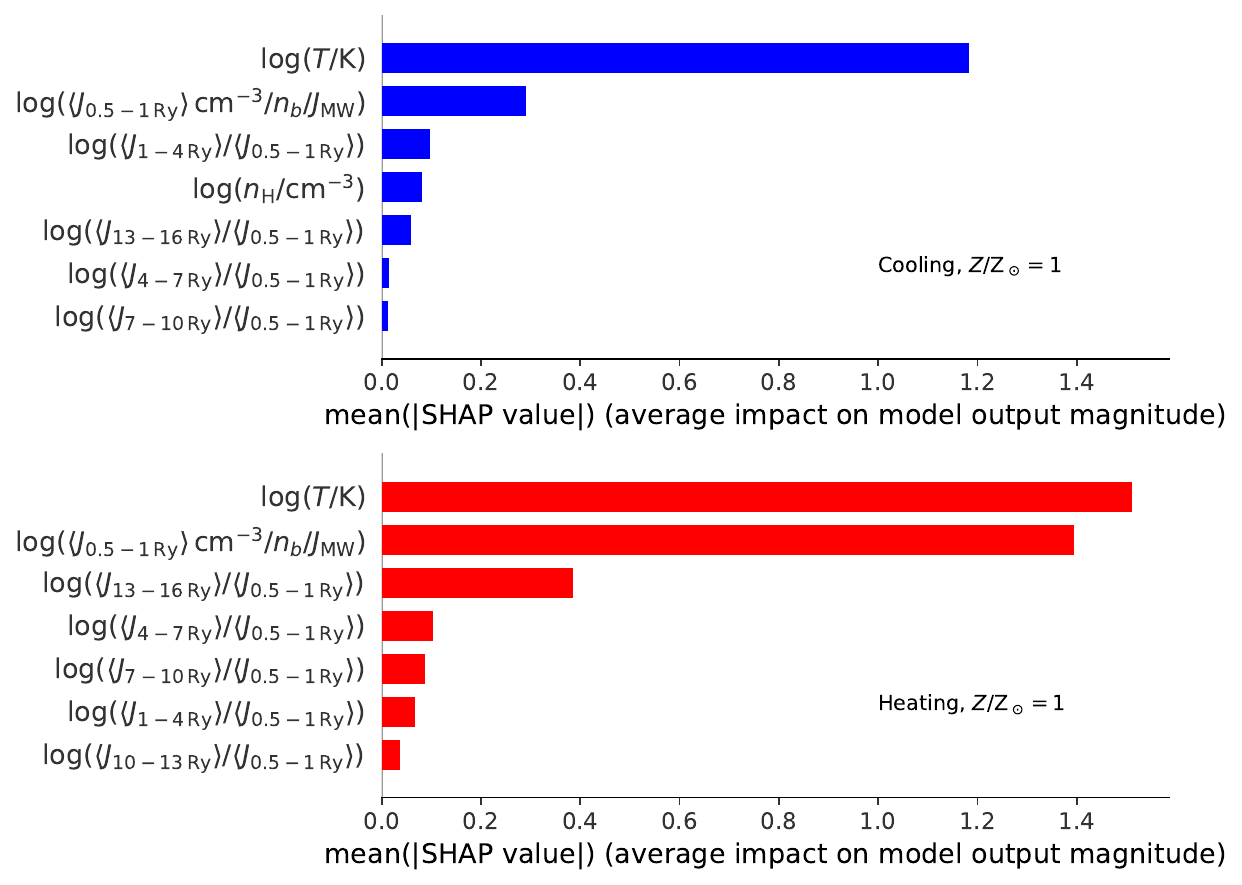}
    \caption{Mean absolute SHAP importance on 500 randomly selected points in the test set (20\% of the training grid withheld from model training) for XGBoost models trained using 6 bins (see leftmost column of Table~\ref{tab:features}).  Models for the cooling function (top panel) and heating function (bottom panel) at $Z/\mathrm{Z}_\odot = 1$ are shown. Only the 7 features with the largest mean absolute SHAP values for each model are shown.  At $Z/\mathrm{Z}_\odot = 1$, the gas temperature and overall amplitude of the radiation field are most important.}
    \label{fig:shap_values}
\end{figure*}

We use the Python package \texttt{shap} \citep{lundberg18, lundberg20}, which incorporates functions to approximate SHAP values for tree-based models like XGBoost, to calculate SHAP values for predictions on the 20\% test set from the 10 models trained on the 80\% training subset. We randomly select 500 points from the 20\% test set for each model to use for SHAP value calculations.  Since SHAP values can be either positive or negative (i.e.\ a model prediction can be either increased or decreased, given the value of a specific feature), we consider the \textit{mean absolute SHAP value} over the 500 test points for each feature. As an example of this, Fig.~\ref{fig:shap_values} shows the 7 features with the largest mean absolute SHAP values for cooling and heating function models at $Z/\mathrm{Z}_\odot = 1$ (trained using 6 bins, as described in the leftmost column of Table~\ref{tab:features}). 

As seen in Fig.~\ref{fig:shap_values}, the temperature feature $\log{(T/\mathrm{K})}$ and the $0.5-1 \, \mathrm{Ry}$ bin that includes the overall radiation field amplitude, $\log{(\langle J_{0.5-1 \mathrm{Ry}} \rangle \, \mathrm{cm}^{-3} /n_b/J_\mathrm{MW})}$, are the two most important features for all 10 models.  The density feature, $\log{(n_\mathrm{H}/\mathrm{cm}^{-3})}$, tends to have relatively low mean absolute SHAP values, especially for the heating function. 

\begin{figure*}
    \centering
    \includegraphics[width = 0.8\textwidth]{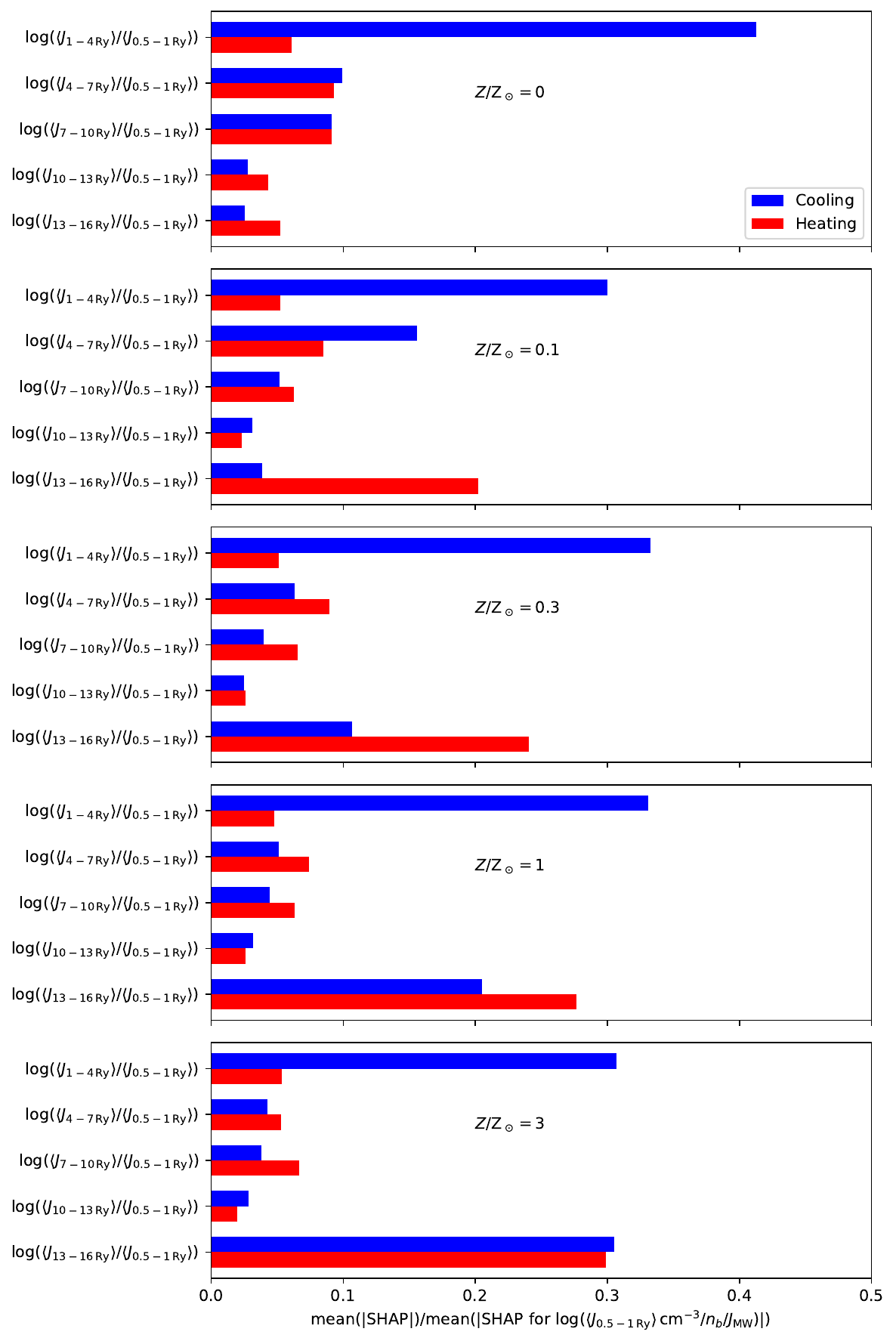}
    \caption{The mean absolute SHAP values from (e.g. Fig.~\ref{fig:shap_values}) for $\log{(\langle J_{\nu_a - \nu_b} \rangle/\langle J_{0.5-1 \, \mathrm{Ry} } \rangle)}$, divided by that for $\langle J_{0.5 - 1 \, \mathrm{Ry}} \rangle)$.  For each feature, the values are shown for the cooling function (above, blue) and the heating function (below, red).  The 5 panels are for models trained at different metallicities, from $Z/\mathrm{Z}_\odot = 0$ at the top to $Z/\mathrm{Z}_\odot = 3$ at the bottom.  Either the $1-4$ or $13-16 \, \mathrm{Ry}$ bins (top and bottom bars in each panel) are the most important for both cooling and heating at each metallicity.}
    \label{fig:scaled_shap}
\end{figure*}

In order to examine the SHAP values for the scaled radiation field bin features more closely, we divide the mean absolute SHAP values for each $\log{(\langle J_{\nu_a - \nu_b} \rangle/\langle J_{0.5-1 \, \mathrm{Ry} } \rangle)}$ feature by the corresponding mean absolute SHAP values for $\log{(\langle J_{0.5-1 \mathrm{Ry}} \rangle \, \mathrm{cm}^{-3} /n_b/J_\mathrm{MW})}$. These scaled SHAP values are shown in Fig.~\ref{fig:scaled_shap} for all 10 models described in Section~\ref{method:ml_methods} above.

Fig.~\ref{fig:scaled_shap} shows that the $13-16 \, \mathrm{Ry}$ bin has high importance \textit{relative to the other radiation field bins} for the heating function at metallicities $Z/\mathrm{Z}_\odot > 0$.  Other bins have similarly low importance for the heating function. The $1-4 \, \mathrm{Ry}$ bin has high relative importance for the cooling function at all metallicities.  The $13-16 \, \mathrm{Ry}$ bin has similar relative importance to the $1-4 \, \mathrm{Ry}$ bin at the highest metallicities $(Z/\mathrm{Z}_\odot = 1, 3)$.  Overall, the $1-4 \, \mathrm{Ry}$ and $13-16 \, \mathrm{Ry}$ are the most important scaled bins. \postsubmit{For $Z/\mathrm{Z}_\odot > 0.1$, the cooling function for $10^4 \lesssim T/\mathrm{K} \lesssim 10^7$ is dominated by high ionization states of heavy elements like carbon, oxygen, neon, and iron \citep{wiersma09}. The relevant ionization thresholds are $\gtrsim 13 \, \mathrm{Ry}$, and so are sampled by the $13-16 \, \mathrm{Ry}$ bin. For example, \ion{Fe}{9} becomes an important coolant in highly photoionized gas \citep[e.g.][]{cohen00, cantalupo10} and has an ionization potential of $\sim 12 \, \mathrm{Ry}$ \citep{NIST_ASD}.}

Motivated by these observations, we also train an additional 10 models (cooling and heating at each metallicity) using only the two scaled bins $1-4 \, \mathrm{Ry}$ and $13-16 \, \mathrm{Ry}$.  These models are trained using exactly the same pipeline as described in Section~\ref{method:ml_methods}. We label these models as `3 bins', since they incorporate 3 radiation field bins ($0.5-1, 1-4,$ and $13-16 \, \mathrm{Ry}$), in contrast to the `6 bins' models described in Section~\ref{method:ml_methods} (see Table~\ref{tab:features} for a comparison between the features used in these models).

Fig.~\ref{fig:scaled_shap} shows that $1-4$ and $13-16 \, \mathrm{Ry}$ are the most important scaled bins at all metallicities. Photoionization of \ion{H}{1} and \ion{He}{1} by photons with energies greater than $1 \, \mathrm{Ry}$ are vital contributions to the heating function \citep{ferland98}.  So, we train additional heating function models using the $1-4 \, \mathrm{Ry}$ bin to include the overall radiation field amplitude $J_0$, and scale the intensity in the $13-16 \, \mathrm{Ry}$ bin by that value.  We label these models as `2 bins', since only the $1-4$ and $13-16 \, \mathrm{Ry}$ bins are utilized.  The features used are described in corresponding column of Table~\ref{tab:features}.  Again, these models are trained using the pipeline of Section~\ref{method:ml_methods}. 

For comparison with these models, we also consider two sets of models trained with photoionization rates $P_j$ (see Equation~(\ref{eq:q_j_def})) as described in \citet{robinson24}. First, we use models trained with $P_j$ for $j=\mathrm{LW}$, \ion{H}{1}, \ion{He}{1}, and \ion{C}{6} (the same rates used in the interpolation table of \citet{gnedin_hollon12}), which we label `4 rates'. As demonstrated in \citet{robinson24}, these models perform comparably well to models trained with other sets of 4 photoionization rates, as well as larger sets of photoionization rates. We also train new models with `3 rates' ($P_j$ for $j=\mathrm{LW}$, \ion{H}{1}, and \ion{C}{6}) to compare with our `3 bins' models described above. Note that the choices of rates and bins for these models are entirely analogous.  The radiation responsible for photodissocation of molecular hydrogen lies in the $0.5-1 \, \mathrm{Ry}$ bin, as described above.  The $1-4 \, \mathrm{Ry}$ contains \ion{H}{1}-ionizing radiation, and the $13-16 \, \mathrm{Ry}$ bin and \ion{C}{6} photoionization rate both sample $J_\nu$ at X-ray energies. The features for these `4 rates' and `3 rates' models are described in Table~\ref{tab:features}.

\section{Results}
\label{sec:results}

As seen in \citet{robinson24}, XGBoost models trained on the data described in Section~\ref{method:training_data} occasionally make very large `catastrophic errors'. To track these, the primary metric we use to evaluate the performance of our models is the \textit{distribution} of all errors in a given sample, rather than the MSE. Specifically, we define the error distribution function
\begin{equation}
    \label{eq:error_cdf}
    P(> \Delta \log \mathcal{F}) = \frac{\mathrm{Points \, with \, error \, above \,} \Delta \log \mathcal{F}}{\mathrm{Total \, number \, of \, points}},
\end{equation}
where, once again, $\mathcal{F} = \Gamma$ or $\Lambda$.  

\begin{figure*}
    \centering
    \includegraphics[width = 0.9\textwidth]{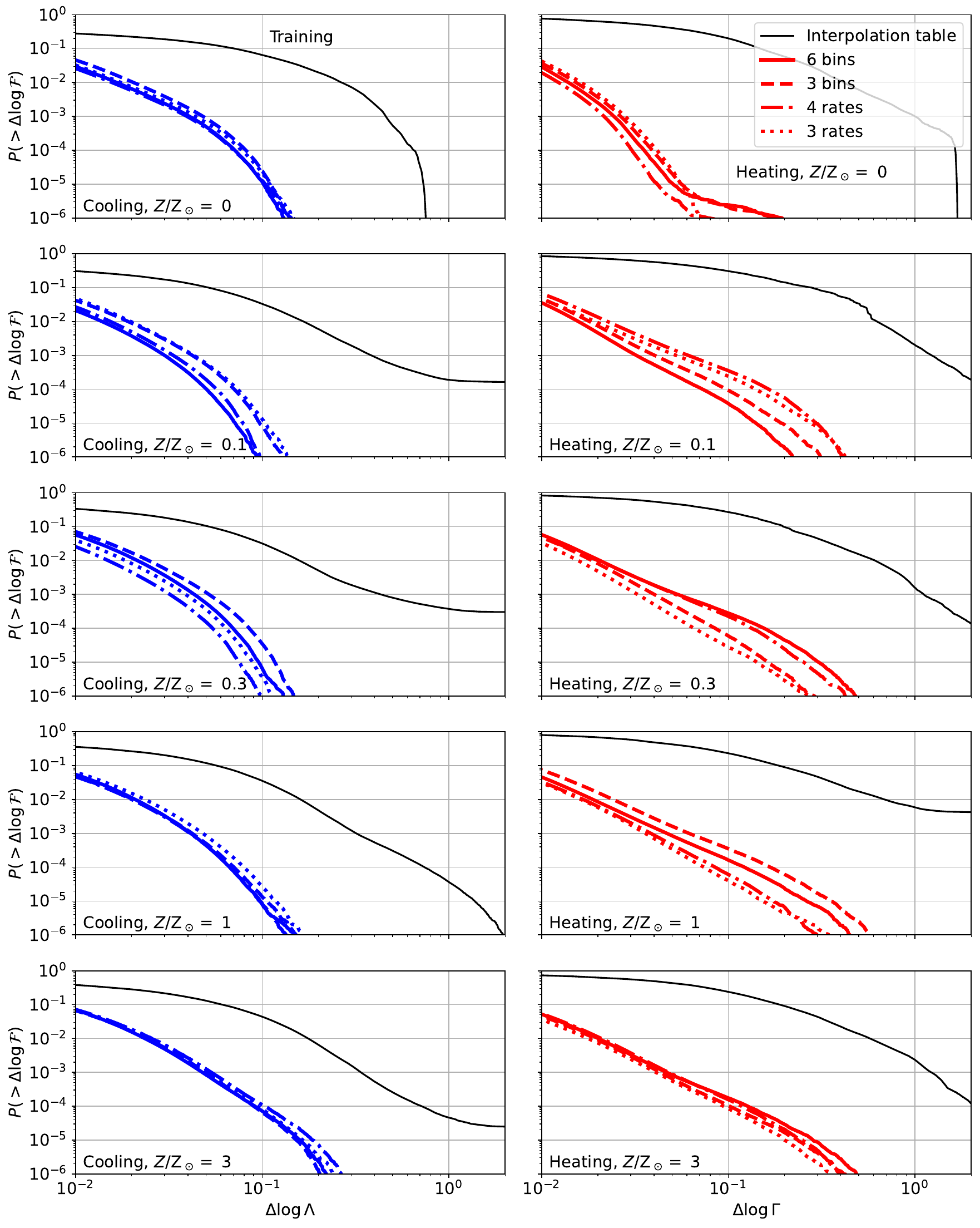}
    \caption{Error distributions for the interpolation table of \citet{gnedin_hollon12} (thin solid lines) and our XGBoost models trained with 6 bins (thick solid lines, described in Section~\ref{method:ml_methods}), 3 bins (thick dashed lines, described in Section~\ref{method:shap}), 4 rates (thick dash-dotted lines, described in Section~\ref{method:shap}), and 3 rates (thick dotted lines, described in Section~\ref{method:shap}) for models trained and evaluated on the entire training grid.  See Table~\ref{tab:features} for the features included in each model.  Error distributions for the cooling function are shown in the left column, with the heating function in the right column. All models are trained to minimize the MSE on the training set.  Our XGBoost models all have similar performance on the training data, and significantly outperform an interpolation table.}
    \label{fig:cdf_train}
\end{figure*}

We examine error distributions for all the cooling and heating function models described in Table~\ref{tab:features}, at each metallicity value.  We show the error distribution on the entire input data for models trained on the entire input data, which assesses how well the models are able to fit this data, in Fig.~\ref{fig:cdf_train}.  The error distributions on the 20\% withheld test set for models trained on the 80\% training subset, which assesses how predictive the models are, are discussed in Appendix~\ref{ap:mse_comp}. We compare the performance of our XGBoost models on the entire training grid with that of the interpolation table from \citet[][see Fig.~\ref{fig:cdf_train}]{gnedin_hollon12}.  The corresponding MSE values are shown in Appendix~\ref{ap:mse_comp}.  We also compare the MSEs for 3 bin and 2 bin heating function models in Appendix~\ref{ap:two_bin_models}.  Since both the training and test set MSEs are about two orders of magnitude higher for the 2 bin models than the 3 bin models, we do not include the 2 bin heating function models in Fig.~\ref{fig:cdf_train}.

For both cooling and heating functions, and at each metallicity value, the performance of the models trained with 3 or 6 radiation field bin features and 3 or 4 photoionization rates is comparable, and much better than the interpolation table of \citet{gnedin_hollon12}. This can be seen through the similarity of the thick solid, dashed, dash-dotted, and dotted training data error distribution curves in Fig.~\ref{fig:cdf_train}. Note that the interpolation table of \citet{gnedin_hollon12} uses a quadratic Taylor expansion in metallicity. The errors for the interpolation table are dominated by errors in the metallicity interpolation \citep[see][]{robinson24}.  Because of this, our XGBoost models are able to outperform the interpolation table at fixed metallicity. However, the \citet{gnedin_hollon12} interpolation scheme is more general, and works at all metallicities (within the range $0 \leq Z/\mathrm{Z}_\odot \leq 3$ of the training data in Table~\ref{table:grid_params}). 

The MSEs for 3 and 6 bin and 3 and 4 rate models differ only by factors of order unity (on either the entire training grid, or the test subset). The MSEs on the test set are always larger than the corresponding training set MSEs by a factor of order unity (again, see Appendix~\ref{ap:mse_comp}).  The heating function models using 2 radiation field bin features have much larger MSEs than the 3 bin models for both the training and test sets (see Appendix~\ref{ap:two_bin_models}, where this is explored further).  

For the heating function at $Z/\mathrm{Z}_\odot=0.3$, and both cooling and heating at $Z/\mathrm{Z}_\odot = 3$, we observe that the training data MSEs for the 3 bin models are rather counterintuively \textit{smaller} than the training data MSEs for the corresponding 6 bin models (see Appendix~\ref{ap:mse_comp}). Note that, since the dependence of the cooling functions and heating function on radiation field bin features $\log{(\langle J_{\nu_a - \nu_b} \rangle/\langle J_{0.5-1 \, \mathrm{Ry} } \rangle)}$ is non-linear, training on a finite set of data results only in the \emph{approximately} optimal dependence of the cooling and heating functions on these binned radiation field features. In other words, while the XGBoost training procedure minimizes the MSE \textit{on a given set of training data}, it is entirely possible that the model performance could be improved with additional training data (that better samples the dependence of the cooling or heating function at the given metallicity on the features being used).  For the models in question, our training data table may not be large enough to take full advantage of the the extra information provided by the additional 3 scaled radiation field bin features in the 6 bin models. Also, for a given number of trees and maximum tree depth (see Appendix~\ref{ap:hyperparams}), the predictions are likely to be more precise for a model with \textit{fewer} features. 
This is because each feature will be used to split the input data more often than in a model with a greater number of features. 

\section{Conclusions and Discussion}
\label{sec:concl}
In this work, we use machine learning to analyze the cooling and heating functions of gas in the presence of a generalized radiation field.  The radiation field contains a synthesized stellar spectrum, a quasar-like power law, and absorption by neutral hydrogen and helium. We describe the radiation field with averaged intensities in discretized photon energy bins. We use the $0.5-1 \, \mathrm{Ry}$ bin to encode the overall amplitude of the radiation field and scale all other binned intensities by that value. Then, we examine the correlations between such binned intensities. We train machine learning models to predict the true (i.e. Cloudy-computed) cooling and heating functions of the gas at 5 fixed metallicities with the 6 least correlated radiation field intensity bins, and the subsequently identified 3 most important radiation field bins, to predict cooling and heating functions.  We also compare these models to \citet{robinson24} models trained using 3 and 4 photoionization rates. The main conclusions from this analysis are:

\begin{itemize}
    \item For our radiation field model (Equation~(\ref{eq:rad_field})), very little additional information is included above a photon energy of $16 \, \mathrm{Ry}$ (see Fig.~\ref{fig:bin_corr}). This is because, at such high energies, the radiation field (up to a factor of the overall amplitude $J_0$) is strongly dependent on the quasar-like power law index $\alpha$, and only weakly dependent on the other parameters ($f_Q, \tau_0$, see Fig.~\ref{fig:rf_examples}).  
    \item For XGBoost models trained with 6 radiation field intensity bins, the $0.5-1 \, \mathrm{Ry}$ bin (containing the overall radiation field amplitude) and temperature are always the two most important features by mean absolute SHAP values (e.g., see Fig.~\ref{fig:shap_values}).
    \item Among the scaled radiation field bin intensities, the $13-16 \, \mathrm{Ry}$ bin has the largest mean absolute SHAP value for heating function models at $Z/\mathrm{Z}_\odot > 0$.  The $13-16$ and/or $1-4 \, \mathrm{Ry}$ bins have significantly larger mean absolute SHAP values than the other bins for cooling function models at all metallicities (see Fig.~\ref{fig:scaled_shap}).
    \item The mean squared errors (MSEs) and error distributions (on both the training data, and a 20\% subset withheld from training) are very similar for models trained with 3 and 6 radiation field bins, and also for 3 and 4 photoionization rates. \postsubmit{These errors are $\gtrsim 10$ times smaller than for the interpolation table of \citet{gnedin_hollon12} applied to the same data.}  However, the MSEs are significantly worse for the heating function models trained with only 2 bins (see Fig.~\ref{fig:cdf_train} and Appendix~\ref{ap:mse_comp}).
    \item We conclude that just 3 photon energy bins ($0.5-1, 1-4$, and $13-16 \, \mathrm{Ry}$) or 3 photoionization rates ($P_j$ for $j=\mathrm{LW}$, \ion{H}{1}, and \ion{C}{6}) are sufficient to capture the dependence of gas cooling and heating functions on an incident radiation field with a spectrum given by Equation~(\ref{eq:rad_field}).
\end{itemize}

\postsubmit{We find that XGBoost models trained using both photoionization rates and energy bins to describe the incident radiation field are able to interpolate Cloudy calculations of gas cooling and heating functions more accurately than the interpolation table of \citet{gnedin_hollon12} at fixed metallicity. The `4 rates' XGBoost model is slower to evaluate than the interpolation table while utilizing the same input parameters. However, adding new radiation field parameters to an interpolation table means increasing the dimension and data size of the table. Because each tree in an XGBoost model only uses a given subset of the available features, the data size of XGBoost models (with the same hyperparameters) scales only weakly with the number of features. This allows us to construct models (such as the `6 bins' models presented here) which incorporate more radiation field parameters than the \citet{gnedin_hollon12} interpolation table without significantly increasing the evaluation time from a model with fewer features.}

Given that our radiation field model (see Equation~(\ref{eq:rad_field})) has 4 free parameters, it is surprising that 3 radiation field features are enough to capture the dependence of cooling and heating functions on that radiation field. In particular, the high energy power-law tail of the radiation spectrum is set by the power-law slope \postsubmit{$\alpha$} and the amplitude \postsubmit{$J_0$}, and these two parameters cannot be captured by the single value in the  $13-16 \, \mathrm{Ry}$ band. 

Our conclusion that 3 energy bins are enough to accurately capture the cooling and heating functions with XGBoost relies on our specific choices of energy bins. \postsubmit{We expect the most important energy bins to be robust to changing the atomic data set or using a different photoionization code. The overall amplitude of the radiation field and H and He ionization radiation in the 1-4 Ry bin will always be important, and cooling at $Z/\mathrm{Z}_\odot > 0.1$ is dominated by metal-line cooling \citep[e.g.][]{wiersma09}, ensuring that radiation in the quasar-dominated part of the spectrum capable of accessing the highest ionization states of metals such as carbon, oxygen, and iron will have a significant effect.} However, the machine learning pipeline used in this work (Section~\ref{method:ml_methods}) could be used to train new cooling and heating function models with \textit{any} set of radiation field energy bins, and in particular a set used in a given radiative transfer simulation.

The radiation field sampling in Equation~(\ref{eq:rf_int}) can be done with any frequency bounds $\nu_a$ and $\nu_b$, and any weight functions $w_j(\nu)$.  In this work, we have only consider two cases, constant weights $w_j(\nu)=\mathrm{const}$ and photoionization cross sections $w_j(\nu)\propto\sigma_j(\nu)$.  As shown in Fig.~\ref{fig:cdf_train} and Appendix~\ref{ap:mse_comp}, 3 such radiation field samples are sufficient to describe the cooling and heating functions reasonably accurately in both cases.  It is of course possible that with different, more optimal weighting functions $w_j(\nu)$, one could find a set of only 2 radiation field samples sufficient to specify the gas cooling and heating functions or a set of 3 samples with significantly smaller errors than the samples presented here. However, finding such weighting functions is a non-trivial exercise.

We have also assumed the radiation field spectrum is well-described by Equation~(\ref{eq:rad_field}). While this is fairly general, it of course does not include every possible radiation field.  For example, it only include absorption by neutral hydrogen and helium, and not other elements or ions.  Our machine learning framework is capable of handling a more complex radiation field model, and future work could involve running more Cloudy models to expand or add more dimensions to the training data in Table~\ref{table:grid_params}. \postsubmit{The Cloudy calculations in the training data used here include only \textit{atomic} cooling and heating processes. Additional processes such as cooling and heating from dust grains and cosmic ray heating could be incorporated by adding additional dimensions (e.g.\ describing the dust abundance and cosmic ray ionization rate) into the training data in Table~\ref{table:grid_params}. These processes might introduce additional important radiation field energy bins, perhaps including photons capable of ejecting an electron from dust grains and heating the gas. Non-solar element abundances would also change the cooling and heating functions from the values in the training data. Abundances for a few particularly important elements could also be included as additional dimensions of the training data in Table~\ref{table:grid_params}. While incorporating non-solar abundances would require training new XGBoost models, we expect our qualitative conclusions about the most significant energy bins and photoionization rates to be robust as long as there are non-trivial amounts of any important metal coolants whose higher ionization states are excited by photons in the quasar-dominated region of the radiation field spectrum described by the $13-16 \, \mathrm{Ry}$ energy bin.}

\section*{Acknowledgments}
DR and CA acknowledge support from the Leinweber Foundation.  CA acknowledges support from DOE grant DE-SC009193. This manuscript has been co-authored by Fermi Research Alliance, LLC under Contract No. DE-AC02-07CH11359 with the U.S. Department of Energy, Office of Science, Office of High Energy Physics.  This research was also supported in part through computational resources and services provided by Advanced Research Computing (ARC), a division of Information and Technology Services (ITS) at the University of Michigan, Ann Arbor, in particular the Great Lakes cluster and the U-M Research Computing Package.

This work utilizes many Python packages, including \texttt{xgboost} \citep{chen_guestrin16}, \texttt{shap} \citet{lundberg18, lundberg20}, \texttt{numpy} \citep{numpy}, \texttt{pandas} \citep{pandas}, \texttt{matplotlib} \citep{matplotlib}, \texttt{seaborn} \citep{seaborn}, \texttt{scikit-learn} \citep{scikit-learn}, and \texttt{scikit-optimize} \citep{scikit-optimize}. The code pipeline to train and evaluate the machine learning models presented in this paper can be found at \url{https://github.com/davidbrobins/ml\_chf}.

\bibliographystyle{mnras}
\bibliography{oja_template}

\begin{appendix}

\begin{figure*}
    \centering
    \includegraphics[width = \textwidth]{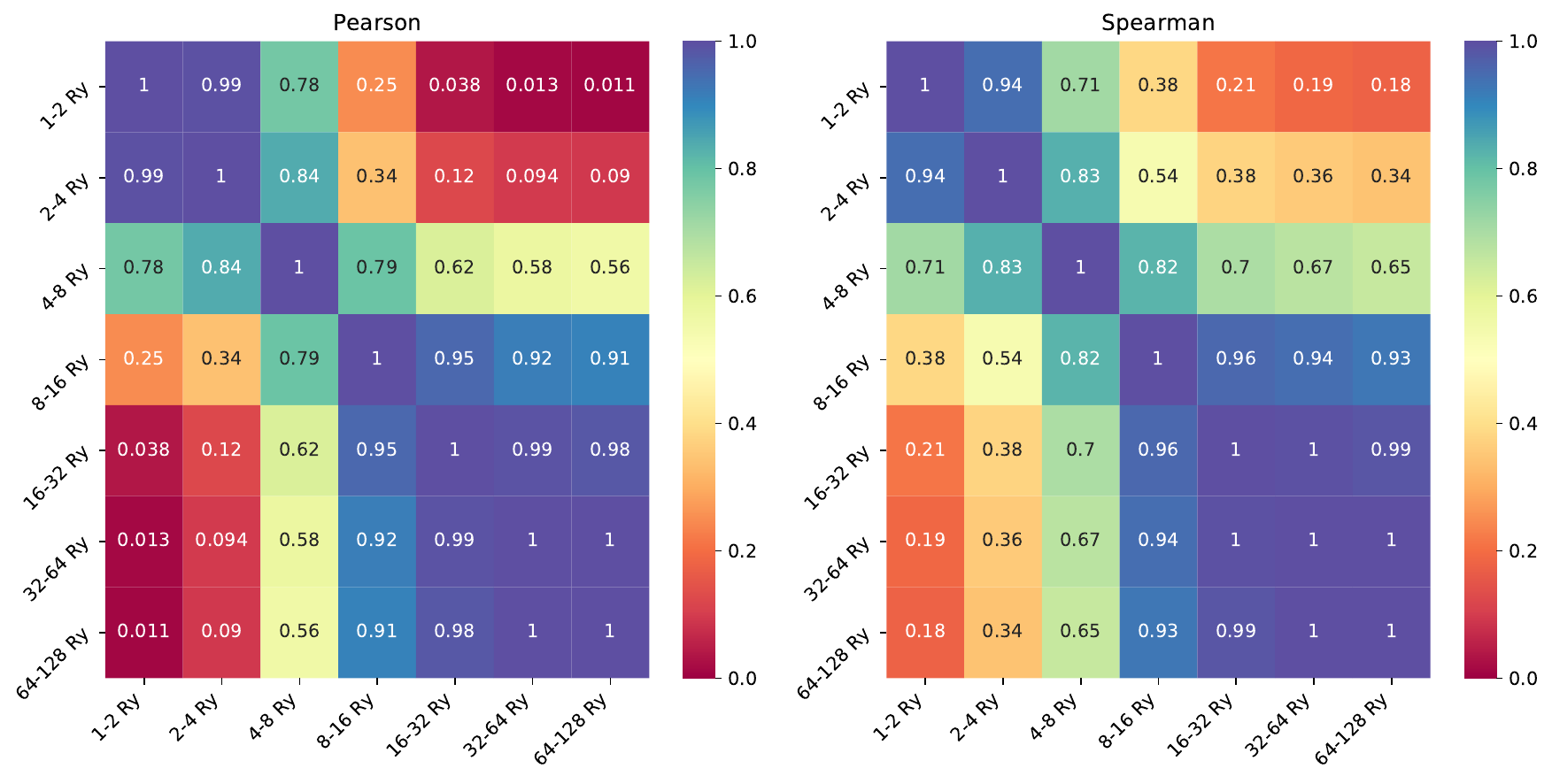}
    \caption{The absolute value of the Pearson (left panel) and Spearman (right panel) correlation coefficients between binned radiation field intensities $\log{(\langle J_{\nu_a - \nu_b} \rangle/\langle J_{0.5-1 \, \mathrm Ry} \rangle)}$ for the values in Table~\ref{table:grid_params}, with logarithmic bin spacing. \vspace{0.5cm}}
    \label{fig:log_spaced_bin_corrs}
\end{figure*}

\section{Logarithmic bin spacing}
\label{ap:log_spaced_bins}
In addition to the bin edges with linear spacing shown in Fig.~\ref{fig:bin_corr}, we also consider the correlations of $\log{(\langle J_{\nu_a - \nu_b} \rangle/\langle J_{0.5-1 \, \mathrm{Ry}} \rangle)}$ for bins with logarithmic spacing, from $1-2 \, \mathrm{Ry}$ up to $1024-2048 \, \mathrm{Ry}$.  The Pearson and Spearman correlation matrices for the first 7 of theses bins are shown in Fig.~\ref{fig:log_spaced_bin_corrs} (the remaining five bins are all highly correlated with the $64-128 \, \mathrm{Ry}$ bin.

From Fig.~\ref{fig:log_spaced_bin_corrs}, we see that the $1-2$ and $2-4 \, \mathrm{Ry}$ bins have large correlations with each others for both Pearson ($0.99$) and Spearman ($0.94$). Thus, it is reasonable to combine these into a single $1-4 \, \mathrm{Ry}$, as we do in Section~\ref{methods:rf_bins}.

The correlation matrices for both linear (Fig.~\ref{fig:bin_corr}) and logarithmic (Fig.~\ref{fig:log_spaced_bin_corrs}) bins show that all bins above $16 \, \mathrm{Ry}$ are highly correlated with each other.  As discussed in Section~\ref{methods:rf_bins}, this is unsurprising because the radiation field given by Equation~(\ref{eq:rad_field}) is dominated by the quasar-like power law at these energies (see Fig.~\ref{fig:rf_examples}).  This power law depends on only one parameter, the index $\alpha$. The logarithmic bin spacing used in Fig.~\ref{fig:log_spaced_bin_corrs} results in only 4 bins below $16 \, \mathrm{Ry}$, while the linear spacing we use in Section~\ref{methods:rf_bins} gives 5 bins below $16 \, \mathrm{Ry}$ (exactly the bins used for our 6 bin models).

\section{Hyperparameter tuning}
\label{ap:hyperparams}

In the XGBoost hyperparameter validation step described in Section~\ref{method:ml_methods}, we optimize 5 XGBoost parameters:\footnote{These, as well as other XGBoost hyperparameters, are described in more detail in the code documentation: \url{https://xgboost.readthedocs.io/en/stable/index.html}}
\begin{itemize}
    \item \texttt{min\_child\_weight}: minimum `weight' required for each tree leaf.  Larger values result in simpler models. Default value: 1 (cannot be negative).
    \item \texttt{subsample}: the fraction of the training data points used to train each tree. Default value: 1 (cannot be negative or larger than 1).
    \item \texttt{colsample\_bytree}: the fraction of features used to train each tree. Default value: 1 (cannot be negative or larger than 1).
    \item \texttt{gamma}: the minimum improvement in model performance required to split a node. Larger values result in simpler models. Default value: 0 (cannot be negative).
    \item \texttt{eta}: the learning rate.  Default value: 0.3 (cannot be negative or larger than 1).
\end{itemize}
Following \citet{robinson24}, we fix the maximum depth (number of nodes to reach the leaves of the tree) to $\texttt{max\_depth}=20$ and the number of trees to $\texttt{n\_estimators}=100$.  

To optimize \texttt{min\_child\_weight}, \texttt{subsample}, \texttt{colsample\_bytree}, \texttt{gamma}, and \texttt{eta}, we use 5-fold cross-validation.  That is, we split the validation subset (as described in Section~\ref{method:ml_methods}, this is 8\% of the input data) into 5 disjoint subsets.  For each of these 5 `folds', we train a model on the remaining 4 folds and evaluate it on the fold in question.  The performance of each model is the average MSE across each of the 5 folds. To minimize this value, we perform a Bayesian search, with the following priors:
\begin{itemize}
    \item \texttt{min\_child\_weight}: log-uniform prior from 0.1 to 2. 
    \item \texttt{subsample}: uniform prior from 0.6 to 1.
    \item \texttt{colsample\_bytree}: uniform prior from 0.6 to 1.
    \item \texttt{gamma}: uniform prior from 0 to 1.
    \item \texttt{eta}: log-uniform prior from 0.03 and 0.3.
\end{itemize}
This entire procedure is implemented with the function \texttt{BayesSearchCV} in the \texttt{scikit-optimize} package \citep{scikit-optimize}.

\section{Model error comparisons}
\label{ap:mse_comp}
\begin{figure*}
    \centering
    \includegraphics[width = 0.9\textwidth]{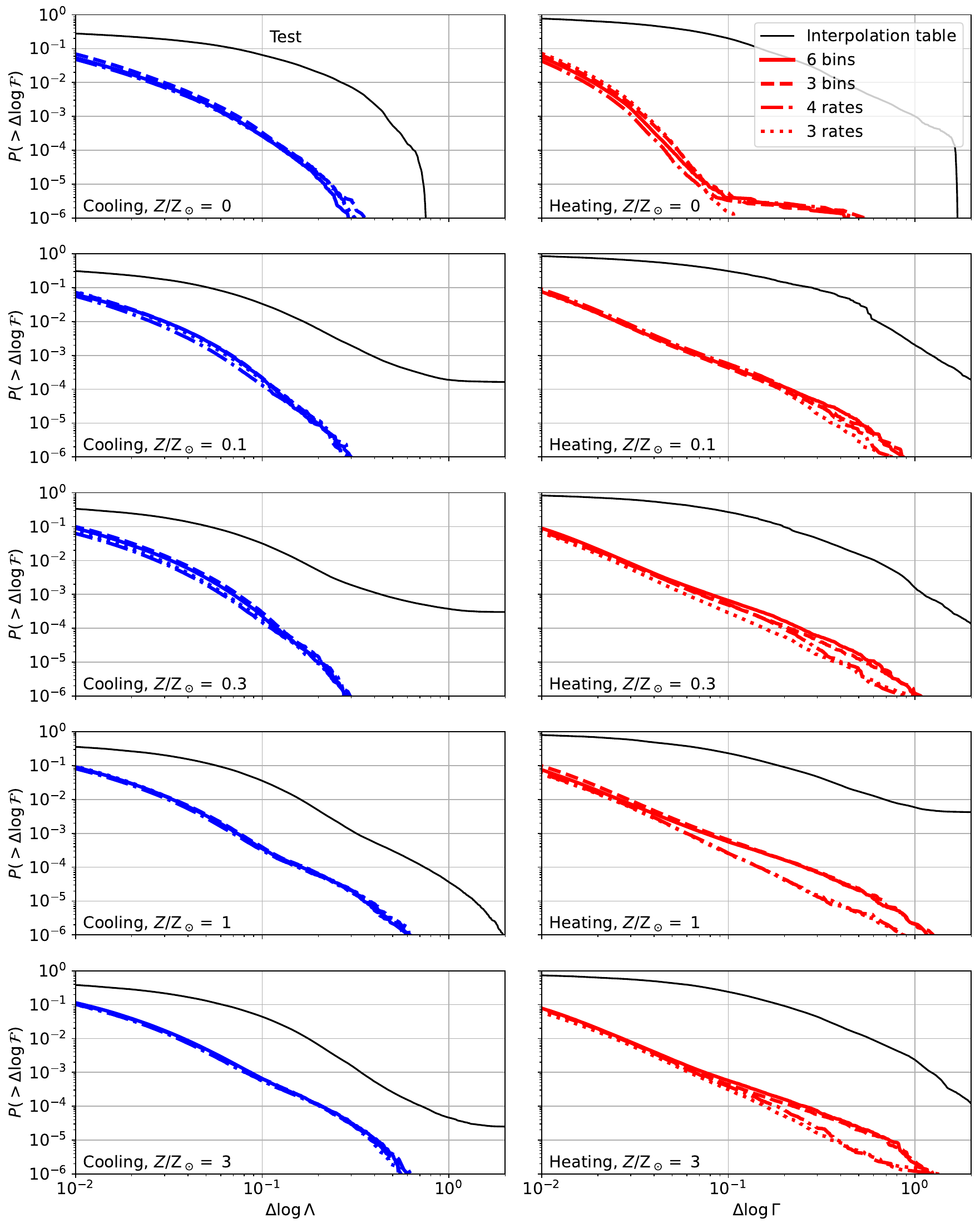}
    \caption{Error distributions for our XGBoost models trained with 6 bins (solid lines, described in Section~\ref{method:ml_methods}), 3 bins (dashed lines, described in Section~\ref{method:shap}), 4 rates (dash-dotted lines, described in Section~\ref{method:shap}), and 3 rates (dotted lines, described in Section~\ref{method:shap}) for models trained on 80\% of the training grid and evaluated on the withheld 20\% test subset (see Section~\ref{method:ml_methods}).  See Table~\ref{tab:features} for the features included in each model.  The error distributions for the interpolation table of \citet{gnedin_hollon12} on the entire training grid are overplotted as thin solid lines.  Distributions for the cooling function are shown in the left column, with the heating function in the right column. All models are trained to minimize the MSE on the training set.  All our XGBoost models have comparable performance on the test subset.}
    \label{fig:cdf_test}
\end{figure*}

In this section, we present further details on the performance on our XGBoost models.  The error distributions (see Equation~(\ref{eq:error_cdf})) on the 20\% withheld test set for models trained on the 80\% training subset are shown in Fig.~\ref{fig:cdf_test}.  We also overplot the error distributions for the interpolation table of \citet{gnedin_hollon12} on the entire data grid (described in Section~\ref{method:training_data}).  Similarly to Fig.~\ref{fig:cdf_train}, the error distributions for all 4 XGBoost models shown in Fig.~\ref{fig:cdf_test} are similar, and much better than the interpolation table of \citet{gnedin_hollon12}.

The MSEs for the XGBoost models in Fig.~\ref{fig:cdf_train} and Fig.~\ref{fig:cdf_test} are shown for cooling function models in Table~\ref{tab:cf_mse} and for heating function models in Table~\ref{tab:hf_mse}.  As discussed in Section~\ref{sec:results}, all of the MSEs in each row of Table~\ref{tab:cf_mse} or Table~\ref{tab:hf_mse} are of the same order of magnitude.

\begingroup 
    \setlength{\tabcolsep}{6pt} 
    \renewcommand{\arraystretch}{1.5} 
    \begin{table*}
        \centering
        \begin{tabular}{rllllllll}
        Cooling & \multicolumn{2}{l}{6 bins} &   \multicolumn{2}{l}{3 bins} & \multicolumn{2}{l}{4 rates} & \multicolumn{2}{l}{3 rates} \\
        & Training & Test & Training & Test & Training & Test & Training & Test \\
        \hline \hline
        $Z/\mathrm{Z}_\odot = 0$ & $1.65 \times 10^{-5}$ & $4.05 \times 10^{-5}$ & $2.52 \times 10^{-5}$ & $5.18 \times 10^{-5}$ & $1.49 \times 10^{-5}$ & $3.76 \times 10^{-5}$ & $1.89 \times 10^{-5}$ & $3.85 \times 10^{-5}$ \\
        $Z/\mathrm{Z}_\odot = 0.1$ & $1.10 \times 10^{-5}$ & $4.62 \times 10^{-5}$ & $2.15 \times 10^{-5}$ & $4.94 \times 10^{-5}$ & $1.36 \times 10^{-5}$ & $3.58 \times 10^{-5}$ & $2.41 \times 10^{-5}$ & $4.77 \times 10^{-5}$ \\
        $Z/\mathrm{Z}_\odot = 0.3$ & $2.72 \times 10^{-5}$ & $5.83 \times 10^{-5}$ & $3.63 \times 10^{-5}$ & $6.69 \times 10^{-5}$ & $1.28 \times 10^{-5}$ & $4.04 \times 10^{-5}$ & $1.96 \times 10^{-5}$ & $4.56 \times 10^{-5}$ \\
        $Z/\mathrm{Z}_\odot = 1$ & $2.58 \times 10^{-5}$ & $6.63 \times 10^{-5}$ & $2.39 \times 10^{-5}$ & $6.66 \times 10^{-5}$ & $2.29 \times 10^{-5}$ & $5.94 \times 10^{-5}$ & $3.25 \times 10^{-5}$ & $6.82 \times 10^{-5}$ \\
        $Z/\mathrm{Z}_\odot = 3$ & $3.48 \times 10^{-5}$ & $9.10 \times 10^{-5}$ & $3.46 \times 10^{-5}$ & $8.64 \times 10^{-5}$ & $3.95 \times 10^{-5}$ & $8.06 \times 10^{-5}$ & $3.84 \times 10^{-5}$ & $8.02 \times 10^{-5}$ \\
        \hline \hline
        \end{tabular}
        \caption{Mean squared errors (MSEs, see Equation~(\ref{eq:MSE})) for the cooling function XGBoost models in the left column of Fig.~\ref{fig:cdf_train} and Fig.~\ref{fig:cdf_test} (see Table~\ref{tab:features}).  All MSEs have similar orders of magnitude.}
        \label{tab:cf_mse}
    \end{table*}
\endgroup

\begingroup 
    \setlength{\tabcolsep}{6pt} 
    \renewcommand{\arraystretch}{1.5} 
    \begin{table*}
        \centering
        \begin{tabular}{rllllllll}
            Heating & \multicolumn{2}{l}{6 bins} & \multicolumn{2}{l}{3 bins} & \multicolumn{2}{l}{4 rates} & \multicolumn{2}{l}{3 rates} \\
            & Training & Test & Training & Test & Training & Test & Training & Test \\
            \hline \hline
            $Z/\mathrm{Z}_\odot = 0$ & $1.53 \times 10^{-5}$ & $2.59 \times 10^{-5}$ & $1.81 \times 10^{-5}$ & $2.93 \times 10^{-5}$ & $1.10 \times 10^{-5}$ & $2.08 \times 10^{-5}$ & $1.97 \times 10^{-5}$ & $3.15 \times 10^{-5}$ \\
            $Z/\mathrm{Z}_\odot = 0.1$ & $1.94 \times 10^{-5}$ & $6.16 \times 10^{-5}$ & $2.64 \times 10^{-5}$ & $5.88 \times 10^{-5}$ & $4.51 \times 10^{-5}$ & $6.74 \times 10^{-5}$ & $3.25 \times 10^{-5}$ & $5.46 \times 10^{-5}$ \\
            $Z/\mathrm{Z}_\odot = 0.3$ & $3.83 \times 10^{-5}$ & $7.80 \times 10^{-5}$ & $2.58 \times 10^{-5}$ & $6.57 \times 10^{-5}$ & $3.22 \times 10^{-5}$ & $5.79 \times 10^{-5}$ & $1.79 \times 10^{-5}$ & $4.65 \times 10^{-5}$ \\
            $Z/\mathrm{Z}_\odot = 1$ & $2.97 \times 10^{-5}$ & $7.33 \times 10^{-5}$ & $4.99 \times 10^{-5}$ & $8.69 \times 10^{-5}$ & $1.96 \times 10^{-5}$ & $4.38 \times 10^{-5}$ & $1.77 \times 10^{-5}$ & $4.17 \times 10^{-5}$ \\
            $Z/\mathrm{Z}_\odot = 3$ & $3.23 \times 10^{-5}$ & $7.68 \times 10^{-5}$ & $2.75 \times 10^{-5}$ & $6.79 \times 10^{-5}$ & $3.28 \times 10^{-5}$ & $5.82 \times 10^{-5}$ & $2.17 \times 10^{-5}$ & $4.61 \times 10^{-5}$ \\
            \hline \hline
        \end{tabular}
        \caption{Mean squared errors (MSEs, see Equation~(\ref{eq:MSE})) for the heating function XGBoost models in the right column of Fig.~\ref{fig:cdf_train} and Fig.~\ref{fig:cdf_test} (see Table~\ref{tab:features}).  All MSEs have similar orders of magnitude.}
        \label{tab:hf_mse}
    \end{table*}
\endgroup

\section{The performance of 2 bin models}
\label{ap:two_bin_models}
We compare the MSEs of our 2 bin and 3 bin heating function models in Table~\ref{tab:2_bin_hf_mse}.  Models trained with only 2 bins have MSEs that are about 2 orders of magnitude higher than the corresponding 3 bin models, both for models evaluated on the entire training grid and on a 20\% withheld test set (see Table~\ref{tab:features} for the specific features used in each model). 

\begingroup 
    \setlength{\tabcolsep}{6pt} 
    \renewcommand{\arraystretch}{1.5} 
    \begin{table*}
        \centering
        \begin{tabular}{rllll}
            Heating & \multicolumn{2}{l}{3 bins} & \multicolumn{2}{l}{2 bins} \\
            & Training & Test & Training & Test \\
            \hline \hline
            $Z/\mathrm{Z}_\odot = 0$ & $1.81 \times 10^{-5}$ & $2.93 \times 10^{-5}$ & $2.01 \times 10^{-3}$ & $2.79 \times 10^{-3}$ \\
            $Z/\mathrm{Z}_\odot = 0.1$ & $2.64 \times 10^{-5}$ & $5.88 \times 10^{-5}$ & $4.95 \times 10^{-3}$ & $6.82 \times 10^{-3}$ \\
            $Z/\mathrm{Z}_\odot = 0.3$ & $2.58 \times 10^{-5}$ & $6.57 \times 10^{-5}$ & $5.90 \times 10^{-3}$ & $9.35 \times 10^{-3}$ \\
            $Z/\mathrm{Z}_\odot = 1$ & $4.99 \times 10^{-5}$ & $8.69 \times 10^{-5}$ & $7.03 \times 10^{-3}$ & $1.09 \times 10^{-2}$ \\
            $Z/\mathrm{Z}_\odot = 3$ & $2.75 \times 10^{-5}$ & $6.79 \times 10^{-5}$ & $7.68 \times 10^{-3}$ & $1.18 \times 10^{-2}$ \\
            \hline \hline
        \end{tabular}
        \caption{Mean squared errors (MSEs, see Equation~(\ref{eq:MSE})) for 3 bin and 2 bin heating function models (see Table~\ref{tab:features}).  The 2 bin model is systematically $\sim$2 orders of magnitude worse than the 3 bin model.}
        \label{tab:2_bin_hf_mse}
    \end{table*}
\endgroup

To understand why the 2 bin models have such large MSEs, we choose a single point from the training data where the heating function error at $Z/\mathrm{Z}_\odot = 0$ satisfies $\Delta \log \Gamma > 0.8$. The values of each input to the training table for the selected point are shown in Table~\ref{tab:bad_2_bin_point}.

\begingroup 
    \setlength{\tabcolsep}{6pt} 
    \renewcommand{\arraystretch}{1.5} 
    \begin{table}
        \centering
        \begin{tabular}{ r l }  
            Parameter & Values \\
            \hline \hline     
            $\log{(T/\mathrm{K})}$ & $1$ \\
            $\log{(n_\mathrm{H}/\mathrm{cm}^{-3})}$ & $-6$ \\   
            $Z/\mathrm{Z}_\odot$ & $0$ \\
            \hline 
            $\log{(J_0 \, \mathrm{cm}^{-3}/n_b/J_\mathrm{MW})}$ & $-5$ \\
            $\log{(f_Q)}$ & $1$ \\
            $\log{(\tau_0)}$ & $2$ \\
            $\alpha$ & $0$ \\
            \hline \hline
        \end{tabular}  
        \caption{Parameters for a point in the training grid (see Table~\ref{table:grid_params}) with $\Delta \log \Gamma > 0.8$ for a 2 bin heating function model at $Z/\mathrm{Z}_\odot = 0$.}
        \label{tab:bad_2_bin_point}
    \end{table}
\endgroup

\begin{figure}
    \centering
    \includegraphics[width = 0.8\textwidth]{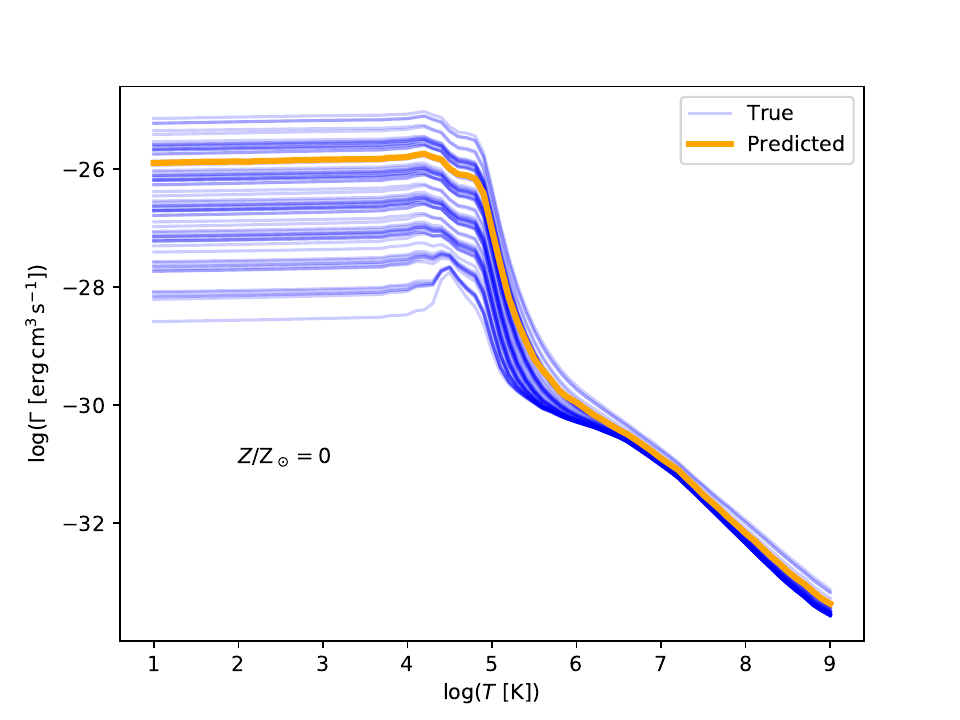}
    \caption{The predicted heating function for the training grid point described in Table~\ref{tab:bad_2_bin_point} (thick orange curve), and the true heating functions for every training grid point with the same density, and binned radiation field features (see Table~\ref{tab:features}) within 10\% of the values for the point in Table~\ref{tab:bad_2_bin_point} (thin light blue curves).}
    \label{fig:high_err_2_bins}
\end{figure}

Then we select all other points in the training grid with the same density, and values for the two binned radiation field features, $\log{(\langle J_{1-4 \, \mathrm{Ry}} \rangle \, \mathrm{cm}^{-3}/n_b/J_\mathrm{MW})}$ and $\log{(\langle J_{13-16 \, \mathrm{Ry}} \rangle / \langle J_{1-4 \, \mathrm{Ry}} \rangle)}$ within 10\% of the values of the point described in Table~\ref{tab:bad_2_bin_point} \textit{after scaling to the interval} $0-1$ ($0.512886$ and $0.306554$, respectively). In Fig.~\ref{fig:high_err_2_bins}, we plot the true heating function (i.e. as a function of temperature) from Cloudy for each such point, along with the predicted heating function for the point in Table~\ref{tab:2_bin_hf_mse}.  

While the true heating functions in Fig.~\ref{fig:high_err_2_bins} are fairly similar for $T \gsim 10^{6.5} \, \mathrm{K}$, they span almost 3 orders of magnitude for $T \lsim 10^4 \, \mathrm{K}$.  So, any single prediction would by necessity have $\Delta \log \Gamma \sim 1$ at $T \lsim 10^4 \, \mathrm{K}$ for some of the true heating functions shown in Fig.~\ref{fig:high_err_2_bins}. This indicates that points with similar values of the radiation field features for our 2 bin models (see Table~\ref{tab:features}) can have very different heating functions, suggesting that these two radiation field features are insufficient to capture the behavior of the heating function.

\begin{figure}
    \centering
    \includegraphics[width = 0.9\textwidth]{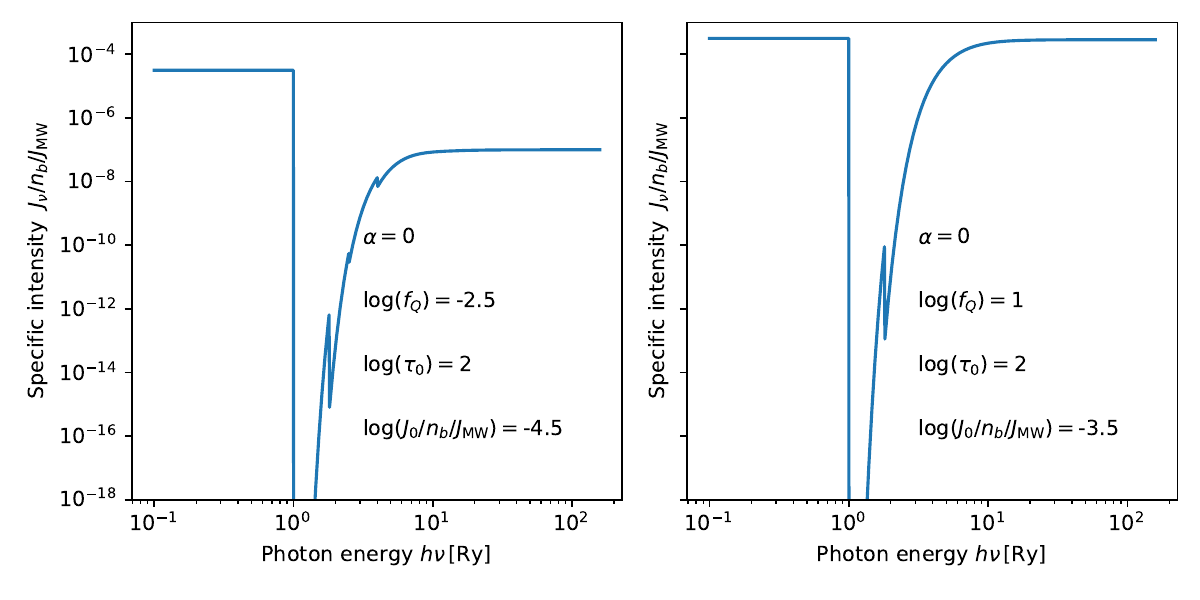}
    \caption{The radiation field specific intensity $J_\nu$ (Equation~(\ref{eq:rad_field})) for the points from Fig.~\ref{fig:high_err_2_bins} with the lowest (left panel) and highest (right panel) heating function values at $\log(T/\mathrm{K})=1$.}
    \label{fig:extreme_rfs}
\end{figure}

To illustrate this, we plot the radiation field specific intensity $J_\nu$ for the points with the highest and lowest heating function values at $\log(T/\mathrm{K})=1$ from Fig.~\ref{fig:high_err_2_bins}. These spectra are shown in Fig.~\ref{fig:extreme_rfs}.

The two spectra in Fig.~\ref{fig:extreme_rfs} have the same values of optical depth scaling $\tau_0 = 100$ and quasar-like power law index $\alpha = 0$, but very different values of the fractional contribution of quasars, $f_Q$.  The value of $J_\nu$ plateaus at energies above about $10 \, \mathrm{Ry}$ in both spectra. In the right panel, the high value of $f_Q$ means that $J_\nu$ is as large above $10 \, \mathrm{Ry}$ as it is below $1 \, \mathrm{Ry}$.  However, in the left panel, the value of $J_\nu$ above $10 \, \mathrm{Ry}$ is almost 3 orders of magnitude smaller than that below $1 \, \mathrm{Ry}$.  This difference would be captured by the radiation field features of our 3 and 6 bin models (see Table~\ref{tab:features}), which utilize the $0.5-1 \, \mathrm{Ry}$ bin to scale all other bins. However, it is not captured by scaling with the $1-4 \, \mathrm{Ry}$ bin used in our 2 bin heating function models. 

The intensity in the $0.5-1 \, \mathrm{Ry}$ bin (used to include $J_0$ and for scaling all other bins in our 3 and 6 bin models) has no dependence on the optical depth parameter $\tau_0$, because only photons with energies greater than $1 \, \mathrm{Ry}$ can ionize \ion{H}{1} or \ion{He}{1} (see Equation~(\ref{eq:opt_depth})).  On the other hand, the intensity in the $1-4 \, \mathrm{Ry}$ is \textit{strongly} sensitive to the value of $\tau_0$ (see Fig.~\ref{fig:rf_examples}). The $13-16 \, \mathrm{Ry}$ bin, should only have weak dependence on $\tau_0$.  Scaling by the $1-4 \, \mathrm{Ry}$ bin (as is done in our 2 bin models) introduces additional dependence on $\tau_0$. This creates a degeneracy in the radiation field $J_\nu$ implied by our 2 bin model features (see Table~\ref{tab:features}).  This is the same degeneracy seen in the spread of the heating function at low temperatures in Fig.~\ref{fig:high_err_2_bins}.

\end{appendix}

\end{document}